\documentclass[%
 reprint,
 amsmath,amssymb,
 aps,
]{revtex4-2}

\usepackage{tabularx}
\usepackage{gensymb,units,textcomp}
\usepackage{graphicx}% Include figure files
\usepackage{dcolumn}% Align table columns on decimal point
\usepackage{bm}% bold math
\usepackage{xcolor}
\usepackage{subcaption}
\begin{document}

\preprint{}

\title{Exploring extreme magnetization phenomena in directly-driven imploding cylindrical targets}

\author{C. A. Walsh$^{1}$}
\author{R. Florido$^{2}$}
\author{M. Bailly-Grandvaux$^{3}$}
\author{F. Suzuki-Vidal$^{4}$}
\author{J. P. Chittenden$^{4}$}
\author{A. J. Crilly$^{4}$}
\author{M. A. Gigosos$^{5}$}
\author{R. C. Mancini$^{6}$}
\author{G. P\'erez-Callejo$^{7}$}
\author{C. Vlachos$^{7}$}
\author{C. McGuffey$^{8}$}
\author{F. N. Beg$^{8}$}
\author{J. J. Santos$^{7}$}

\affiliation{$^{1}$Lawrence Livermore National Laboratory}
\affiliation{$^{2}$iUNAT - Departamento de F\'isica, Universidad de Las Palmas de Gran Canaria, 35017 Las Palmas de Gran Canaria, Spain}
\affiliation{$^{3}$Center for Energy Research, University of California, San Diego}
\affiliation{$^{4}$Blackett Laboratory, Imperial College London, London SW7 2AZ, UK}
\affiliation{$^{5}$Departamento de F\'isica Te\'orica, At\'omica y \'Optica, Universidad de Valladolid, 47071 Valladolid, Spain}
\affiliation{$^{6}$Department of Physics, University of Nevada, Reno, NV 89557, USA}
\affiliation{$^{7}$Université de Bordeaux-CNRS-CEA, Centre Lasers Intenses et Applications (CELIA),\\ UMR 5107, F-33405 Talence, France}
\affiliation{$^{8}$General Atomics, San Diego}

\date{\today}

\begin{abstract}

This paper uses extended-magnetohydrodynamics (MHD) simulations to explore an extreme magnetized plasma regime realisable by cylindrical implosions on the OMEGA laser facility. This regime is characterized by highly compressed magnetic fields (greater than 10~kT across the fuel), which contain a significant proportion of the implosion energy and induce large electrical currents in the plasma. Parameters governing the different magnetization processes such as Ohmic dissipation and suppression of instabilities by magnetic tension are presented, allowing for optimization of experiments to study specific phenomena. For instance, a dopant added to the target gas-fill can enhance magnetic flux compression while enabling spectroscopic diagnosis of the imploding core. In particular, the use of Ar K-shell spectroscopy is investigated by performing detailed non-LTE atomic kinetics and radiative transfer calculations on the MHD data. Direct measurement of the core electron density and temperature would be possible, allowing for both the impact of magnetization on the final temperature and thermal pressure to be obtained. 
%Compared to an unmagnetized implosion, it is expected that these measurements will show the core temperature increasing and the density decreasing happening in the magnetized scenario due to decreased thermal conduction and the increased conversion of implosion energy into magnetic energy. 
By assuming the magnetic field is frozen into the plasma motion, which is shown to be a good approximation for highly magnetized implosions, spectroscopic diagnosis could be used to estimate which magnetization processes are ruling the implosion dynamics; for example, a relation is given for inferring whether thermally-driven or current-driven transport is dominating. 
\end{abstract}

\maketitle

%---------------------------------------------------------
\section{\label{sec:intro}Introduction}
%---------------------------------------------------------

Magnetic fields are of interest in astrophysics, existing in the Universe on all scales: from kilometers or less for compact neutron stars up to millions of light years for clusters of galaxies. Natural magnetic fields can exceed the laboratory magnetic fields achieved by many orders of magnitude, with white dwarfs sustaining magnetic fields of up to $10^5$~T and the most extreme neutron stars up to $10^9$~T ~\cite{lai2001}. In such compact stars the magnetic fields are strong enough to modify the structure and composition of the star, as well as its radiation properties~\cite{engel2008,murdin2013}.

In the laboratory, laser-driven plasma compression has been used to amplify initial (seed) magnetic fields \cite{hohenberger2012}; to first order the magnetic field is frozen into the plasma as it compresses. Proton deflectometry has provided a direct measurement of these compressed magnetic fields up to $4$~kT in dedicated cylindrical experiments \cite{gotchev2009,knauer2010}. Novel techniques to generate strong magnetostatic fields $\sim1-10$s~T by either capacitor-bank pulsed discharges~\cite{pollock2006,gotchev2009,albertazzi2013,fiksel2015} or intense laser-driven discharges~\cite{fujioka2013,santos2015,law2016,goyon2017,tikhonchuk2017,santos2018} have stimulated a growing interest in laser-driven, high-energy-density (HED) systems with embedded magnetic fields. 

In Inertial Confinement Fusion (ICF), highly compressed magnetic fields have been demonstrated to contribute to increase implosion performance~\cite{chang2011}. The compressed B-field strengths can be enough to magnetize the electron thermal conduction, reducing the energy losses from the hot-spot and resulting in hotter fuel~\cite{perkins2017,walsh2019}. The B-fields can also effectively confine D-T ions and thermonuclear $\alpha$-particles \cite{sio2021}, enhancing collisionality and fusion yield. In addition, large magnetic fields can reduce hydrodynamic instabilities through magnetic tension \cite{chandrasekhar1962,sano2013,srinivasan2013,perkins2017,walsh2019}, although there are also mechanisms that increase perturbation growth with magnetization \cite{walsh2019,walsh2020a}. 

Cylindrical geometry provides a simple testbed for exploring magnetized phenomena in HED plasmas, as magnetic fields can increase asymmetry in spherical implosions \cite{walsh2019,walsh2020a}. In this scheme, the cylinder axis is  parallel to the applied B-field direction. Using this geometry, the MagLIF concept was first proposed at Sandia National Laboratories (SNL) in 2010 ~\cite{slutz2010}. It consists of laser pre-heat of cold fuel inside a Z-pinch cylindrical liner (initially with an axial B-field) which is magnetically compressed by a $\sim20$~MA, 150~ns electrical current. Due to fuel magnetization and pre-heating, MagLIF aims at near-adiabatic stable cylindrical compression with lower implosion velocity ($\sim100$~km/s, instead of $>300$~km/s) and lower convergence ratios than conventional ICF. MagLIF has shown promising results through different experimental campaigns by demonstrating thermonuclear neutron generation at fusion-relevant conditions, a high enough fuel magnetization for $\alpha$-particles trapping~\cite{gomez2014,gomez2015,knapp2015} and reporting first performance scaling studies~\cite{gomez2020}.

A laser-driven approach to MagLIF is being explored at the OMEGA Laser facility~\cite{davies2017,barnak2017,hansen2018,davies2019,hansen2020} to provide a test for MagLIF scaling. The driving laser energy is $1000\times$ less than that delivered on the Z-pinch facility at SNL and therefore the laser targets are $10\times$ smaller than on Z. Yet, OMEGA provides a significantly higher repetition rate and better diagnostic access, facilitating the parametric study of the underlying basic physics. Cylindrical implosions are easier to perform and offer a clear axis for diagnostics, as well as for a laser preheat~\cite{barnak2017} or a fast ignition beam. A recent series of laser-driven experiments~\cite{hansen2020} used the improved MIFEDS~\cite{gotchev2009,fiksel2015} to deliver a seed B-field of nearly 30 T, which was high enough to demonstrate the effects of magnetic pressure decreasing plasma compression. There are prospects to deliver even higher B-field values ($\sim50$~T) within the next few years. %Among the drawbacks, the MIFEDS system may limit the diagnostic access to the imploding plasma and also the large mass of the hardware components constitutes an important --and potentially damaging-- source of debris in the interaction chamber. In this regard, laser-driven coils (LDC) targets also show promise for magnetizing target implosions~\cite{santos2015,law2016,tikhonchuk2017,bailly-grandvaux2018,sakata2018,santos2018}, although the short rise time of the magnetic field is yet to be demonstrated to allow diffusion into the centre of the target before implosion. On the upside, this technique does not need any external electric power source as it is optically driven by laser beams, it provides easier diagnostic access and, relatively to MIFEDS, limits the debris in the interaction chamber thanks to the small mass of the mm-scale LDC targets.   

The design and interpretation of all above experiments and applications strongly rely on magneto-hydrodynamic (MHD) codes. Simulations must include extended-MHD effects, which represent the transport of energy and magnetic flux in a plasma~\cite{walsh2020}. Magnetized plasmas are thought to exhibit complex behavior in the electron population and, above resistive-MHD, extended-MHD additionally accounts for temperature-gradient-driven transport --such as the Nernst term moving magnetic fields down electron temperature gradients-- and electric-current-driven transport --such as the Hall term moving magnetic fields with the flow of charge--. The Nernst term, in particular, is of great interest to the scientific community, with importance in a broad range of experiments \cite{chang2011,gao2015,campbell2020,gomez2020,tubman2021} with expectations of plasma demagnetization \cite{joglekar2016,hill2017,sherlock2020,walsh2020a,walsh2017}. Even with its wide applicability, Nernst advection of magnetic field is yet to be directly measured.

This paper explores the development of extreme magnetization phenomena in directly-driven imploding cylindrical targets using extended-MHD simulations. The starting point for the design is a previous laser-driven implosion on OMEGA (without any fuel pre-heat), which found evidence of significant magnetic pressure during target stagnation~\cite{hansen2020}. Distinct magnetization phenomena are outlined for this high B-field regime, which can be categorised as related to magnetic field amplification, electron magnetization or induced electrical currents; metrics governing which phenomena are significant are summarized in Table \ref{tab:1}. Initial target conditions are varied to transition between different regimes; for example, the concentration of a dopant added to the imploding fuel can be increased to amplify magnetic field compression at the expense of electron magnetization, while enabling a spectroscopic characterization of plasma conditions. In particular, the feasibility of Ar K-shell spectroscopy~\cite{regan2002,welser2007,florido2011,florido2014,nagayama2014,carpenter2020a,carpenter2020b} for diagnosing the core conditions in magnetized cylindrical implosions is demonstrated by  post-processing the extended-MHD simulations and performing non-LTE atomic kinetics and radiative transfer calculations. By using the frozen-in-flow approximation, a spectroscopic diagnosis of core conditions would allow for estimates of the magnetization metrics.

% and perform an exhaustive simulation study to investigate the impact of applying different seed B-field values (from 0~T up to 50~T) on the hydrodynamical behavior of the imploding target and the resulting B-field amplification, electron magnetization or induced electrical currents. We notice that results discussed here are independent of the technique employed to generate the initial B-field and therefore our conclusions are in line with the current status and prospects of both MIFEDS system and LDC targets.  

% begins presenting basic simulation results. The compressing B-field changes the hydrodynamic behavior, producing significant variations in the radial profiles and time-histories of core temperature and density throughout the implosion compared to the non-magnetized case

This paper is organized as follows. Section~\ref{sec:gorgon} briefly summarizes the main features of the Gorgon code used to perform the extended-MHD simulations. Section~\ref{sec:design_basic_results} discusses the platform for magnetized direct-drive cylindrical implosions and the corresponding nominal target design.  Section~\ref{sec:New_Physics} discusses key magnetization phenomena that may occur in a magnetized compressed hot-spot. Non-dimensional parameters to quantify the relative importance of different physical processes for any given implosion are introduced.  Section~\ref{sec:design} uses the metrics defined in Sec.~\ref{sec:New_Physics} to modify the point-of-design to explore specific physics regimes and scenarios; e.g. maximizing the magnetic field strength requires a different design to maximizing the electron magnetization. Synthetic images of time-resolved Ar K-shell spectra are presented in Section \ref{sec:diagnosis} in an effort to mimic potential measurements recorded by a streaked spectrometer. Lastly, the appendix compares synthetic proton radiographs produced with Gorgon to experimental data recorded in the cylindrical implosions described in Refs.~\cite{gotchev2009,knauer2010}, thus validating the basic magnetic transport processes in the code. Sensitivity of the radiographs to magnetic transport by bulk plasma motion and Nernst demagnetization is observed.

%-------------------------------------------------------------------------------
\section{Gorgon extended-MHD code\label{sec:gorgon}} % SECTION II
%-------------------------------------------------------------------------------

The simulations presented in this paper utilize the Gorgon extended-MHD code \cite{ciardi2007,chittenden2004,walsh2017}. Gorgon includes magnetic transport by temperature gradients (such as the Nernst term), as well as by electrical currents (such as the Hall effect), and the analogous terms advecting electron energy \cite{walsh2020}. These processes will be discussed in more detail in Section \ref{sec:New_Physics}. The transport coefficients preserve physical behaviour at low magnetization \cite{sadler2021}, which can give significant differences \cite{walsh2021a} to the coefficients calculated by Epperlein \& Haines \cite{epperlein1986}. The thermal conduction algorithm is specifically designed for anisotropic heat-flow \cite{sharma2007,walsh2018a}. Biermann battery production of magnetic fields is also taken into account \cite{walsh2017,campbell2020}. Importantly for this work, the plasma dynamics include the Lorentz force, resistive diffusion and Ohmic dissipation. 

Gorgon results have been compared with magnetized cylindrical implosion experiments on OMEGA~\cite{gotchev2009,knauer2010} using synthetic proton radiography, matching the peak proton deflection (and therefore the compressed magnetic field) --details are shown in the appendix. This provides a crucial benchmark for the code and lends additional confidence in using the Gorgon code for the experimental design addressed here, where the main differences are in the increased applied magnetic field and more sophisticated diagnostics (presence of a dopant for spectroscopic purposes). Gorgon simulations of pre-magnetized capsules with 3D perturbations have also been used to study how differing magnetization effects impact on short and long wavelength perturbations, as well as the impact of magnetization on stagnation temperatures and yields~\cite{walsh2019,walsh2020a}.

The laser drive is modelled using a ray-trace package and inverse Bremsstrahlung absorption. %The experimental laser pointing is used to drive the capsule.
The simulations in this paper are predominantly 2D, with two demonstrative 3D simulations in Sec.~\ref{sec:New_Physics}. The 2D simulations are in cylindrical geometry $r-z$, such that the axial plasma end losses are captured. 

Gorgon uses a $P\frac{1}{3}$ automatic flux-limiting scheme for the radiation transport \cite{jennings2006}, which captures both the free-streaming and diffusive limits. As Ar-doped D$_2$ is used to fill the targets, 100 energy groups are considered in order to capture the complex dependence of argon opacity and emissivity on radiation energy, for different concentrations of dopant. It is assumed that the dopant only affects the radiative properties of the fuel, with any changes to the equation of state or electron transport coefficients not incorporated.

Thermal and Nernst flux limiters of 0.1 are used throughout this paper. The simulations are run with a radial resolution of $\unit[\frac{1}{2}]{\micro m}$, which is required in order to resolve the small converged hot-spot size, with final hot-spot diameters on the order of $\unit[10-20]{\micro m}$ \cite{hansen2020}. Gradients in the axial direction are much shallower, allowing for a converged resolution of $\unit[2]{\micro m}$.

%---------------------------------------------------------
\section{\label{sec:design_basic_results}Magnetized cylindrical implosions on OMEGA}
%---------------------------------------------------------

%% FIGURE 1 : VISRAD %%
\begin{figure}
	\centering
	\includegraphics[width=0.8\columnwidth]{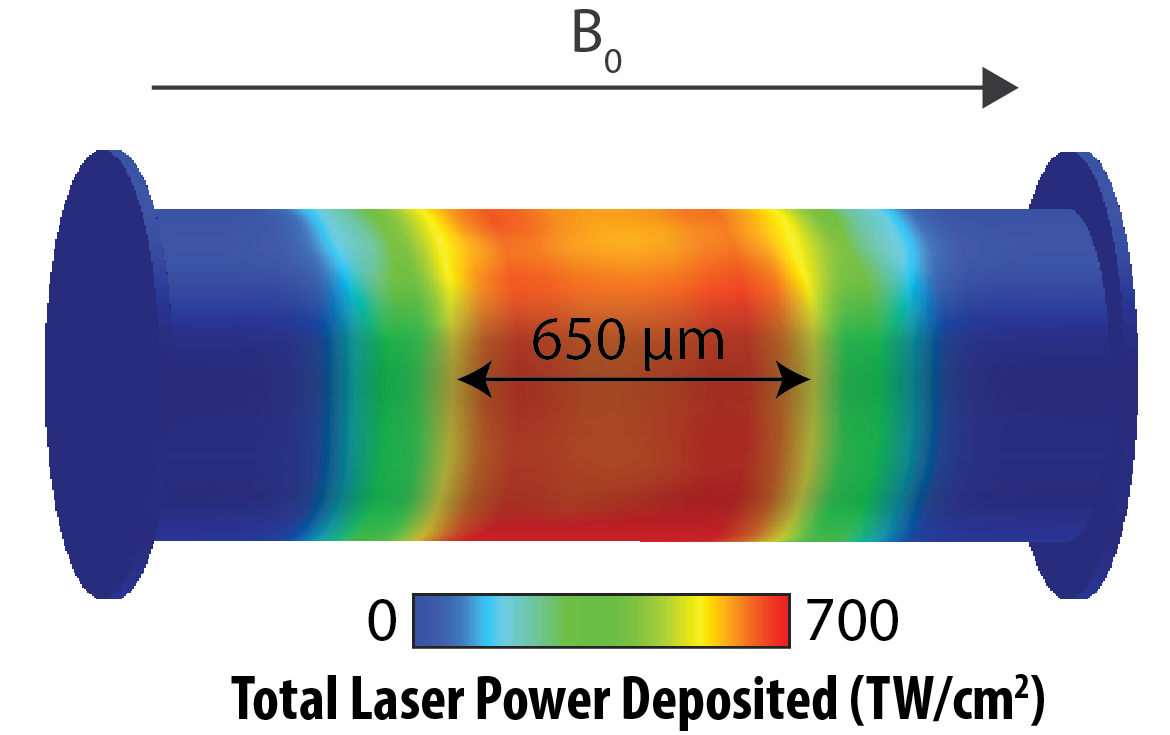}
	\caption{\label{fig:setup_int_pattern} VISRAD image showing the intensity radiation pattern delivered by 40 OMEGA beams on a cylindrical target with an outer radius of \unit[290]{\micro m} (see text for details). A nearly uniform irradiation close to $700$~TW/cm$^2$ is expected on the central $\sim\unit[650]{\micro m}$-length portion along the cylinder axis.}
\end{figure}

%% FIGURE 2 - GORGON 2D SIMS B vs no-B
\begin{figure*}
	\centering
	\includegraphics[scale=0.7]{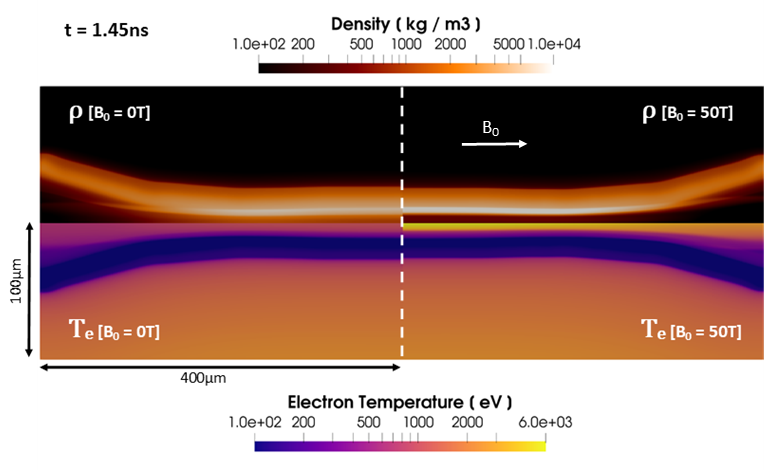}
	\caption{ Density (top) and electron temperature (bottom) for 2D MHD Gorgon simulations of deuterium-filled cylindrical implosions with no magnetic field applied (left) and a 50 T seed magnetic field applied (right). The deuterium fuel has an argon dopant with a concentration of 0.325\% by atom. The data is shown at 1.45~ns, which is approximately 50~ps before the time of peak neutron production.  \label{fig:rho_Te_2D}}
\end{figure*}

While spherical laser-driven implosions are used to obtain high yields, the application of a magnetic field results in anisotropic implosions \cite{walsh2019,walsh2020a}, with the magnetic field normal to the ablator at the pole and in the plane of the ablator at the waist; i.e. a magnetized spherical implosion is intrinsically two-dimensional in its evolution. It was also found that a magnetic field normal to the ablator surface can increase perturbation growth from laser non-uniformities \cite{walsh2020a}. Therefore, by using a cylindrical target with applied field along the axis, these complications are avoided; the magnetic field is allowed to be everywhere in the plane of the ablator, resulting in a one-dimensional implosion geometry before realistic asymmetries are taken into account. Also, while there is more mass compression in a spherical implosion, the magnetic field is not compressed at the poles, so that effective magnetic field compression is similar between the two geometries. 

The nominal target design from recent laser-driven MagLIF implosions (mini-MagLIF campaign) performed at OMEGA is chosen~\cite{hansen2020}, i.e.~targets are cylindrical shells made of parylene-N with a density of $1.11$~g/cm$^3$, with an outer radius of $\unit[290]{\micro m}$  and a wall-thickness of $\unit[17]{\micro m}$. The targets are filled with deuterium at a pressure of $11$~atm, i.e. a mass density of $\unit[1.81]{kg/m^3}$. Targets are imploded by 40 OMEGA beams, delivering a total of \unit[14.5]{kJ} over \unit[1.5]{ns} square-shaped pulses. The laser pointing has been empirically optimized to achieve a uniform implosion velocity over an axial extent greater than \unit[600]{\micro m} \cite{hansen2018,hansen2018a}. Figure~\ref{fig:setup_int_pattern} shows a VISRAD~\cite{macfarlane2003} image with the expected laser power irradiation pattern on target; this illumination is used as an input to the Gorgon simulations.

While the present simulation study originates from the mini-MagLIF platform it is worth noting that, in contrast, the goal here is not to find the regime of maximum fusion performance. The approach is to demonstrate that a regime of extreme magnetization is attainable and investigate how the variation of different target parameters can be used to test complex MHD phenomena. For instance, in this work fuel pre-heating is intentionally not considered as it increases the thermal pressure of the hot-spot at the cost of reducing B-field compressibility. Pre-heating is otherwise essential in the MagLIF design to enhance the neutron yield \cite{gomez2020}. Therefore, the proposed research is distinct --albeit complementary-- to the current mini-MagLIF studies.

As illustration, Fig.~\ref{fig:rho_Te_2D} shows simulation results for this setup using a 0.325\% argon dopant by deuterium atom. On the left is the density and electron temperature without an applied magnetic field, while the right shows the same properties when a 50~T seed B-field is applied. The snapshot is at 1.45 ns, which is 50~ps before the time of peak neutron production. The hot-spot temperature is greatly enhanced by magnetization, as the electron thermal losses are suppressed perpendicular to the applied field. The core hot-spot density is also greatly decreased by magnetization; this is due to two factors. %Firstly, the magnetic field hinders the motion of plasma fuel particles inwards in the radial direction. Secondly, the magnetic pressure in this implosion is significant, so when adding to the thermal pressure, the target compressibility is effectively lowered. 
Firstly, changes in thermal conduction do not change the overall hot-spot thermal pressure \cite{betti2001}. Therefore, an increased temperature by electron magnetization results in a lower density. Secondly, the final total pressure is the sum of thermal and magnetic components; as the magnetic pressure increases, the stagnated thermal pressure decreases, reducing hot-spot compression.

Simulations show that the irradiation pattern of Fig.~\ref{fig:setup_int_pattern} drives an implosion with a central $\sim\unit[400]{\micro m}$-long portion for which core conditions show very little variations in the axial direction. This can be seen in Fig.~\ref{fig:rho_Te_2D} at the corresponding time of 1.45~ns, but the same behaviour is found over the entire implosion evolution.

As the magnetic flux is compressed along with the plasma, the B-field is strongest in the stagnated fuel. The field strength is greater than 10~kT throughout the D$_2$ plasma when an initial 50~T field is applied to the target. While the outer core is denser than its central portion, i.e. the hot-spot, the magnetic field is able to diffuse over the short radial core length-scales, resulting in a relatively uniform magnetic field throughout the fuel. 

\begin{figure}
\centering
\begin{subfigure}[b]{0.45\textwidth}
%\centering
\includegraphics[width=\textwidth]{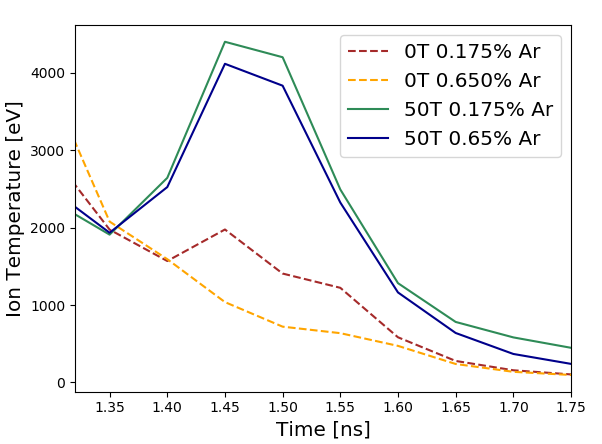}
%\caption{}
%\label{fig:Press_balance}
\end{subfigure}

\begin{subfigure}[b]{0.45\textwidth}
%\centering
\includegraphics[width=\textwidth]{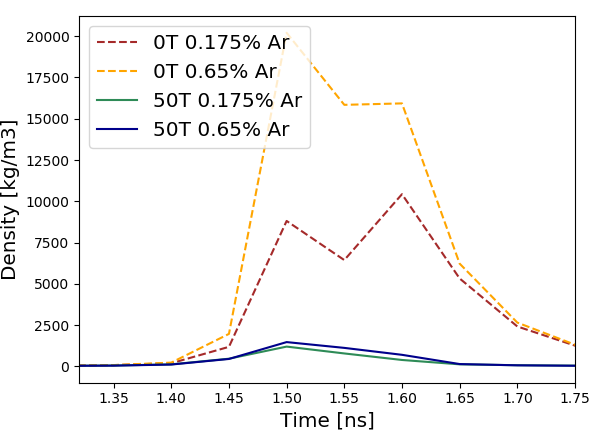}
%\caption{}
%\label{fig:Press_balance}
\label{fig:time_hist}
\end{subfigure}

\caption{Time history of core conditions from 2D Gorgon simulations of cylindrical implosions as a function of different argon dopant concentrations and applied field strength (see Fig.~\ref{fig:rho_Te_2D}). Shown values are calculated as instantaneous neutron-averaged quantities, which results in a weighting towards the hot-spot core.}
\label{fig:time_histories}
\end{figure}

Figure \ref{fig:time_histories} shows the temporal evolution of the implosions by plotting the instantaneous neutron-averaged ion temperature and mass density, e.g. the burn-averaged ion temperature is the ion temperature that a fusing D-D reaction sees (a spatial and temporal average). The burn-average is chosen as it weights the properties to those occurring in the hot-spot, where the extreme magnetization phenomena typically take place. Cases with and without a 50~T seed applied field are shown, as well as for two argon dopant concentrations. Clearly the compressed magnetic field enhances the temperature but significantly decreases the hot-spot density. The increased dopant also results in more radiated energy, lowering the temperature but enhancing the compression of hot-spot mass. When compared to unmagnetized implosions, the compression of the imposed B-field has a significant impact on the conditions of the imploding core, which leads to different regimes of extreme magnetization. This raises the question about how to measure and characterize the corresponding magnetization scenario in an implosion experiment. A detailed assessment about the use of Ar K-shell spectroscopy for this purpose is addressed in Sec.~\ref{sec:diagnosis}.

%\note{[A key difference from the enclosed design to previous experiments is the use of argon dopant in the fuel. Varying the dopant concentration allows for increasing radiative losses, which increases the compressibility of the hot-spot and magnetic field. While high-Z dopant would not be used while aiming for a high fusion yield, here it is used to enhance the magnetic field compression. The radiative losses from the dopant are also used as a diagnostic hot-spot conditions, as the Ar K-shell emission lines (produced at the predicted conditions) are appropriately used for diagnosing the temperature and density of the electron population.]}
 
%% FIG. 3 - TIME HISTORIES, NEUTRON AVG. 

%-------------------------------------------------------------------------------
\section{Magnetization phenomena and associated metrics \label{sec:New_Physics}}
%-------------------------------------------------------------------------------

%\begin{figure}
%		\centering
%		\includegraphics[scale=0.5]{./3D_Pert_Growth.png}\caption{ Density profiles at time of peak neutron production for 3D simulations with no magnetic field applied (left) and 50T applied (right). Along the initial magnetic field direction ($\underline{z}$) the perturbations are suppressed by the large magnetic tension \cite{walsh2019,srinivasan2013,perkins2017}. However, in the $\underline{x}-\underline{y}$ plane the perturbation growth is larger due to suppressed thermal ablative stabilization \cite{walsh2019}.   \label{fig:3D}}
%	\end{figure}

This section outlines the processes of interest in a magnetized implosion. These phenomena can be split into three categories: the effectiveness of magnetic flux compression, magnetization of the electron population, and induced electrical currents. Metrics for the relative importance of each phenomenon are presented, with design changes explored in Sec.~\ref{sec:design}.

Firstly, the process of hot-spot magnetic flux compression is examined. As the work here is primarily interested in the field strength in the compressed hot-spot, discussion of magnetic transport in the laser absorption region is avoided. For an in-depth discussion on this issue, see Refs.~\cite{walsh2020a,davies2015}.

In order to reach a large magnetization regime in the hot-spot, efficient magnetic flux compression is required. If the magnetic field is completely frozen into the plasma motion, and the plasma motion is one dimensional, then the magnetic field in the fuel would exactly follow $B_z = B_0\frac{\rho}{\rho_0}$. Therefore, to quantify how close the magnetic transport is to being frozen into the plasma during the implosion, a \emph{frozen-in-flow factor} is defined:
\begin{equation}
 \Gamma = \frac{|\underline{B}|}{ B_0} \frac{\rho_0}{\rho},\label{eq:frozen_factor}
\end{equation}
where a value of $\Gamma < 1$ indicates that some magnetic flux has been lost from the hot-spot. The primary factors that change the magnetic flux compression from frozen-in-flow are resistive diffusion, Nernst advection of magnetic fields down temperature gradients and axial plasma motion. It will be seen in Sec.~\ref{sec:design} that $\Gamma \approx 1$ for highly magnetized implosions. In some cases the magnetic flux compression actually exceeds the frozen-in-flow approximation.

While $\Gamma$ indicates how much magnetic flux is lost from the hot fuel, maximizing this value does not necessarily help to enhance the magnetic field strength in the hot-spot. For this purpose, the field compression factor is utilized:
\begin{equation}
 \Lambda = \frac{|\underline{B}|}{ B_0}.
\end{equation}
For an implosion with effective magnetic flux compression ($\Gamma = 1$) the field compression factor is simply:
\begin{equation}
\Lambda = \frac{\rho}{\rho_0},\quad \text{if } \Gamma = 1, 
\end{equation}
i.e. the more the plasma is compressed, the more the magnetic field is compressed. 

Once the magnetic field is compressed, the electron population can become magnetized, with each electron taking a curved path between collisions. The Hall Parameter $\omega_e \tau_e$ quantifies the number of orbits a typical electron takes between collisions. While linked to the magnetic field strength, the magnetization of electrons is also dependent on the electron temperature and density \cite{huba2013}:
\begin{equation}
	\omega_e \tau_e \propto \frac{| \underline{B}| T_{e}^{3/2}}{n_e}. \label{eq:Hall}
\end{equation}

Again, using the frozen-in-flow assumption, the Hall parameter can be expressed in terms of the initial fill density of the target as well as the applied field strength:
\begin{equation}
\omega_e \tau_e \propto \frac{B_0}{\rho_0} T_{e}^{3/2},\quad \text{if } \Gamma = 1. \label{eq:Hall2}
\end{equation}

Magnetizing the electron population typically requires much lower magnetic field strengths than magnetic pressure effects \cite{walsh2019}. In magnetized fusion experiments, magnetizing the electron population is a primary goal, with thermal conduction suppression resulting in lower energy losses from the fusing plasma \cite{perkins2013}.  The effect of thermal conduction suppression can be clearly seen in the hot-spot temperature in Fig.~\ref{fig:rho_Te_2D} increasing from approximately 1000~eV to 6000~eV when magnetized (although it will be seen later that a fraction of this increase is due to Ohmic dissipation). In the hot-spot the simulated Hall parameter is in excess of 500, reducing the electron thermal conduction perpendicular to the field lines by a factor of more than $2 \times 10^6$.

The Hall parameter also impacts the transport of magnetic fields by the Nernst term. In an extended-MHD plasma, Nernst can be approximated as the advection of magnetic field with the electron heat-flow \cite{haines1986,walsh2020,walsh2021a}. Therefore, the extreme magnetizations simulated in Fig.~\ref{fig:rho_Te_2D} result in the suppression of Nernst by a factor greater than $2 \times 10^6$, allowing for reduced magnetic flux losses from the hot-spot. 

Now the focus turns to induced electrical currents in magnetized implosions. The current takes the form $\underline{j} = \nabla \times \underline{B}/\mu_0$, where the electric field is assumed to vary slowly. Associated with this current is an electron drift velocity $v_e = -\underline{j}/en_e$. This drift is expected to result in: the transport of thermal and magnetic energy at velocity $v_e$; a dynamically important force; an additional heating term as the finite resistivity of the plasma converts magnetic energy into electron thermal energy. Observing these phenomena experimentally would be an interesting outcome of this platform. 

The Lorentz force is recovered by noting that the electron drift is a movement of electrons in a magnetic field $\underline{f} = (\underline{v}_e \times \underline{B})en_e$; a component of this force is the magnetic pressure. In an unmagnetized target the implosion kinetic energy is converted into stagnated thermal energy, with the plasma retaining some residual kinetic energy depending on implosion symmetry. When a large magnetic field is introduced, some of that implosion kinetic energy converts to magnetic energy instead; the imploding shell decelerates on the magnetic pressure rather than the hot-spot pressure. To characterize this, the ratio of the thermal pressure to the magnetic pressure is typically used:
\begin{equation}
    \beta = \dfrac{P_{\mathrm{th}}}{P_{\mathrm{mag}}} = \frac{2 (P_e + P_i) \mu_0}{|\underline{B}|^2} \approx \frac{2 (n_e T_e + n_i T_i) \mu_0}{|\underline{B}|^2}, \label{eq:mag_pressure}
\end{equation}
where an ideal gas equation of state is assumed for the gas-fill.

For the simulation with a 50 T applied magnetic field in Fig.~\ref{fig:rho_Te_2D}, the magnetic pressure is approximately equal to the thermal pressure ($\beta \approx 1$). A combination of the substantial magnetic pressure and the heat-flow suppression accounts for the lower hot-spot density in the simulation when a 50T magnetic field is applied. 

Equation \ref{eq:mag_pressure} can be recast in terms of the initial applied field and gas fill density by assuming $\Gamma = 1$ (i.e frozen-in-flow). Also incorporating the assumption that the ion and electron temperatures are similar gives:
\begin{equation}
\beta = \frac{2 (1+Z)\rho_0^2 T_e \mu_0}{m_i B_0^2 \rho},\quad \text{if } \Gamma = 1, \label{eq:mag_pressure_rho0}
\end{equation}
where $Z$ is the average ionization of the core plasma ($Z=1$ for deuterium) and $m_i$ is the corresponding ion mass. According to Eq.~\ref{eq:mag_pressure_rho0}, a higher temperature hot-spot enhances the thermal pressure, but more compression of the plasma (and therefore magnetic field) enhances the magnetic pressure contribution.

Implosions dominated by magnetic pressure have previously been indirectly inferred in an experiment similar to the configuration discussed here \cite{hansen2020}. The neutron yield was seen to increase by approximately 50\% going from 0 T to 10 T, primarily due to hot-spot thermal losses being reduced. However, going from 10 T to 27 T decreased the yield by around 50\%.  Simulations attributed the decrease to the magnetic pressure increasing in the hot-spot at the expense of the thermal pressure \cite{hansen2020}. While the decrease in thermal pressure is undesirable for reaching high neutron yields, the large magnetic pressure also results in a highly magnetized regime with large induced electrical currents; these currents are of interest to the research proposed here. 

Simulations anticipate that an initial 50 T field is amplified to greater than $2 \times 10^4$ T. Compressing a magnetic field in excess of $10^4$ T in a hot-spot of around 10 $\mu m$ radius results in a large electric current. For a cylindrical implosion the magnetic field remains predominantly axial, giving a purely azimuthal current that can be approximated as:
\begin{equation}
    j_{\theta} \approx -\frac{B_z}{R_{hs} \mu_0}, \label{eq:current}
\end{equation}
where $R_{hs}$ is the hot-spot radius. The final hot-spot radius can also be related to the compressed density with the convergence relation $R_{hs}/R_0 = \sqrt{\rho_0/\rho}$, i.e. mass conservation assumption. Combining this with the frozen-in-flow assumption:
\begin{equation}
    j_{\theta} \approx - \frac{B_0}{\mu_0 R_0} \Big( \frac{\rho}{\rho_0} \Big)^{3/2},\quad \text{if } \Gamma = 1. \label{eq:current2}
\end{equation}
Clearly, a larger initial applied magnetic field and greater plasma compression result in a larger magnitude induced current. 

The electrical resistance of the plasma leads to diffusion of the magnetic field. The associated drop in magnetic energy results in additional energy into the electron population (through Ohmic dissipation). For the magnetized simulation in Fig.~\ref{fig:rho_Te_2D} this contributes to an enhanced electron temperature that exceeds the ion temperature. When Ohmic dissipation is switched off in the code, the peak electron temperature at bang time drops from 6.0~keV to 4.2~keV.

The resistive diffusion timescale is:
\begin{equation}
    t_{\mathrm{diff}} = \frac{R^2 \mu_0}{\eta}, \label{eq:t_diff}
\end{equation}
where $\eta = \alpha_{\perp}^c m_e / (\tau_{ei} e^2 n_e)$ is the magnetic diffusivity and $\alpha_{\perp}^c$ is the dimensionless resistivity that only depends on plasma magnetization and ionization \cite{sadler2021}. The timescale $t_{\mathrm{diff}}$ defines both the rate of magnetic field diffusion as well as the Ohmic heating. For plasmas with $\beta \approx 1$ and resistive diffusion time-scales  on the order of the hot-spot stagnation time ($t_{\mathrm{diff}} \approx t_{\mathrm{stag}}$), the energy exchange from magnetic to electron thermal energy will be significant. Therefore, an Ohmic heating factor $\Pi$ is defined:
\begin{equation}
    \Pi = \frac{t_{\mathrm{stag}}}{t_{\mathrm{diff}}}\frac{1}{\beta}. \label{eq:Ohmic1}
\end{equation}
Using the convergence relation $R/R_0 = \sqrt{\rho_0/\rho}$ and Eqs.~\ref{eq:mag_pressure_rho0}, \ref{eq:t_diff} and \ref{eq:Ohmic1}, the Ohmic heating factor can be re-written:
\begin{equation}
    \Pi = \frac{B_0}{\rho_0 ^3} \frac{t_{\mathrm{stag}} \rho}{\tau_{ei}T_e} \alpha_{\perp}^c \frac{m_i^2 m_e Z}{2(Z+1) e},\quad \text{if } \Gamma = 1.
\end{equation}
Note that the diffusion is greatest at the edge of the core, where the fuel is more dense and lower in temperature.  

In addition to the resistive diffusion of the magnetic field, the electrical current drives magnetic field advection \cite{braginskii1965,walsh2020}. Primarily, this is the transport of the magnetic field at the electron drift velocity $\underline{v}_e = \underline{j}/en_e$, but with collisional corrections. The advection of magnetic field with the electron drift velocity is often called the Hall term. There is a similar transport of the electron energy at the electron drift velocity, which here will be called the thermal Hall term. The similarities in the advection velocities can be made clear by writing \cite{walsh2020}:
\begin{align}
		&\underline{v}_{jB} &=& &-(1+\delta_{\bot}^c)\frac{\underline{j}_{\bot}}{en_e} +\delta_{\wedge}^c(\frac{\underline{j}}{en_e} \times \underline{\hat{b}})  \label{eq:current_driven1}\\
		&\underline{v}_{jU_e} &=&-(1+\beta_{\parallel}^c)\frac{\underline{j}_{\parallel}}{en_e} &-(1+\beta_{\bot}^c)\frac{\underline{j}_{\bot}}{en_e} +\beta_{\wedge}^c(\frac{\underline{j}}{en_e} \times \underline{\hat{b}}),\label{eq:current_driven2}
\end{align}
where $\underline{v}_{jB}$ is the advection of magnetic field due to electrical currents and $\underline{v}_{jU_e}$ is the advection of electron energy due to electrical currents, $\underline{j}_{\parallel}$ represents the component of the current along the magnetic field, and $\underline{j}_{\bot}$ is the component perpendicular. $\underline{\hat{b}}$ is the magnetic field unit vector. The $\beta_{\wedge}^c$ term is known as the Ettingshausen heat-flow. Note that there is no advection of magnetic field parallel to the magnetic field itself, as this has no effect on the magnetic field. The dimensionless transport coefficients $\delta^c$ and $\beta^c$ \cite{walsh2020} provide collisional corrections to the collisionless current-driven phenomena. The $\delta_{\wedge}$ coefficient calculated from Epperlein \& Haines \cite{epperlein1986} gives erroneous behaviour at low magnetization; updated polynomial fits are given in Ref.~\cite{sadler2021}. Transport driven by electrical currents has been discussed theoretically in pulsed power plasma experiments \cite{seyler2018,chittenden1993} and in low density plasma jets \cite{hamlin2018} but to the best of the authors' knowledge has not been measured experimentally. Current-driven transport has not been shown to be significant in dense laser-driven implosions such as those discussed here. 

The Hall (and thermal Hall) velocities are typically larger than the collisional corrections. An approximate expression for this velocity can be found by using Eq.~\ref{eq:current2}, which assumed a purely axial magnetic field:
\begin{equation}
%\label{eq:v_j} 
v_{j} \approx \frac{B_0}{\mu_0 R_0}  \frac{\rho^{1/2}}{\rho_0^{3/2}} \frac{m_i}{Ze}. \label{eq:hall_rho0}
\end{equation}
The drift velocity is larger for a greater applied magnetic field strength. A lower initial fill density $\rho_0$ also enhances the velocity, as the magnetic field is more easily compressed. Note that these relations must be adapted if studying spherical implosions, as the magnetic field lines are not everywhere perpendicular to the implosion velocity. 

A dimensionless number $\Xi$ was defined in Ref.~\cite{walsh2020} as the ratio of the Nernst velocity to the Hall velocity:
\begin{equation}
    \Xi = \frac{\gamma_{\perp}^c \tau_{ei} T_e e n_e \mu_0}{|\underline{B}|m_e}.
\end{equation}
This ratio is a metric for whether the transport is dominated by temperature gradients or electrical currents. Using burn-averaged quantities the implosion shown in Fig.~\ref{fig:rho_Te_2D} has $\Xi<0.002$ for an applied field of 50 T. While this shows the current-driven transport dominating over the thermal transport, this is mainly due to the extreme suppression of Nernst at such high magnetizations.

For 2D $r-z$ simulations, the induced electrical current is purely in the azimuthal direction $\theta$, which means there is no compression of magnetic field by advection with the current and exactly no change in the electron energy profile. The collisional corrections to the Hall and thermal Hall can be in the simulation plane (such as the Ettingshausen $\beta_{\wedge}^c$ term in Eq.~\ref{eq:current_driven2}), but these are smaller than the collisionless component. With variations in the axial direction, the Hall term can induce a twist in the magnetic field profile. This can be seen by decomposing the change in magnetic field due to an advection velocity directed in $\theta$:
\begin{equation}
\Bigg[ \frac{\partial B_{\theta}}{\partial t}\Bigg]_{\mathrm{advection}} = B_r \frac{\partial v_{j\theta}}{\partial r} - B_z \frac{\partial v_{j\theta}}{\partial z}.
\end{equation}

Noting that the magnetic field is dominated by the axial component, a time-scale for twisting the magnetic field can be obtained:
\begin{equation}
    t_{\mathrm{twist}} = \frac{L_{hs}}{2v_j}, \label{eq:twist}
\end{equation}
where $L_{hs}$ is the axial length over which the hot-spot is uniformly driven. Using Eq.~\ref{eq:current}, Eq.~\ref{eq:twist} becomes:
\begin{equation}
    t_{\mathrm{twist}} = L_{hs} R_{hs} \frac{1}{2} \frac{\rho_0}{B_0} \frac{\mu_0 e Z}{m_i}, \label{eq:twist2}
\end{equation}
i.e. both a small hot-spot radius $R_{hs}$ and length $L_{hs}$ are required for significant magnetic field twisting to occur.  

%\begin{equation}
%    t_{twisting} = L \frac{\mu_0 R_0}{B_0}  \frac{\rho_0^{3/2}}{\rho^{1/2}} \frac{Ze}{n_i}
%\end{equation}

For 2D simulations such as those shown in Fig.~\ref{fig:rho_Te_2D}, the Hall term has been included. Due to the large uniformly-driven length ($L_{hs}\approx \unit[600]{\micro m}$) only marginal twisting of the magnetic field occurs. Over the uniformly-driven central length the field twisting $|B_{\theta}|/|\underline{B}| \leq 0.01$. Towards the edges, where 2D effects are most important $|B_{\theta}|/|\underline{B}| \leq 0.1$.

Once the magnetic field is twisted by the Hall term, a component of the induced current emerges in the 2D $r-z$ simulation plane. The electron energy is then affected by the thermal Hall term. A component of the Lorentz force becomes in the $\theta$ direction, which may result in rotation of the plasma. Simulations where the uniform hot-spot length $L_{hs}$ is reduced have been attempted, but once significant $B_{\theta}$ field emerges, the simulations become unstable. This may be related to the natural instability of Braginskii's MHD description \cite{Bell_2020}. 

The stabilization of perturbation growth by large induced electrical currents is of great interest to magnetized ICF implosions \cite{perkins2017,walsh2019,srinivasan2013}. The cylindrical implosions discussed in this paper could be used as a platform for investigating this behaviour. To demonstrate this, 3D simulations are initialized from 1D once the first shock converges onto the axis of the implosion. At the time of initialization the implosion velocity is artificially modulated by $\pm 1\%$ sinusoidally in both $\theta$ and $z$. A mode 5 perturbation is used in $\theta$, which matches the mode number of the baseline laser asymmetry \cite{hansen2020}. In $z$ a perturbation of 20 $\mu$m wavelength is applied. 

%% FIG. 4 - AREAL DENSITIES - ORIGINAL VERSION (C. WALSH)
% \begin{figure}
%      \centering
%      \begin{subfigure}[b]{0.5\textwidth}
%          \centering
%          \includegraphics[width=1.\textwidth]{rhoR_a}
%          \caption{Areal density variation in the axial ($z$) direction with and without a 50T magnetic field applied.}
%          \label{fig:rhoR_a}
%      \end{subfigure}
%      \begin{subfigure}[b]{0.5\textwidth}
%          \centering
%          \includegraphics[width=1.\textwidth]{rhoR_b}
%          \caption{Areal density variation in the azimuthal ($\theta$) direction with and without a 50T magnetic field applied.}
%          \label{fig:rhoR_b}
%      \end{subfigure}
     
%         \caption{Areal density at time of peak neutron production for 3D Gorgon simulations of cylindrical implosions with artificial velocity perturbations applied. In the axial direction the 50~T seed magnetic field lowers perturbation growth due to magnetic tension stabilization \cite{chandrasekhar1962,srinivasan2013,walsh2019}. However, in the azimuthal direction the perturbation growth is enhanced by magnetization, due to decreased thermal ablative stabilization \cite{walsh2019}. The proposed experimental platform could be used to investigate the trade-off between these two effects for varying perturbation wavelengths.}
%         \label{fig:rhoR}
% \end{figure}

%% FIG. 4 - AREAL DENSITIES - PROPOSED VERSION (FSV)
\begin{figure}
    \centering
    \includegraphics[width=0.45\textwidth]{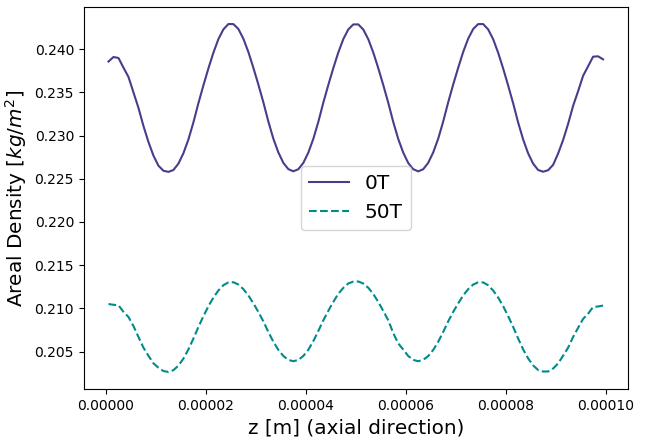}
    \includegraphics[width=0.45\textwidth]{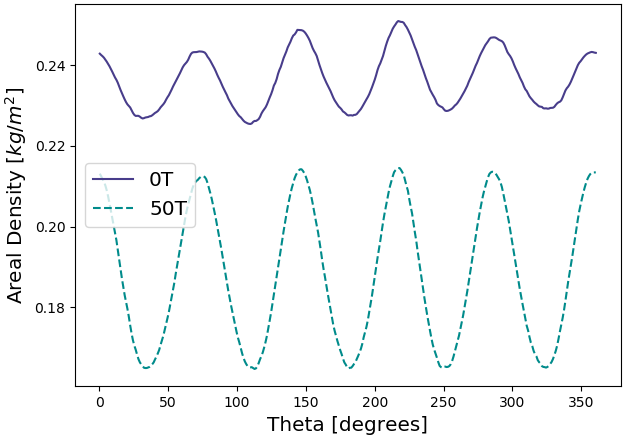}
    \caption{Areal density at time of peak neutron production for 3D Gorgon simulations of cylindrical implosions with an without a 50~T seed magnetic field with artificial velocity perturbations applied. In the axial direction ($z$, top panel) the 50~T seed magnetic field lowers perturbation growth due to magnetic tension stabilization \cite{chandrasekhar1962,srinivasan2013,walsh2019}. However, in the azimuthal direction ($\theta$, bottom panel) the perturbation growth is enhanced by magnetization, due to decreased thermal ablative stabilization \cite{walsh2019}. The proposed experimental platform could be used to investigate the trade-off between these two effects for varying perturbation wavelengths.}
    \label{fig:rhoR}
\end{figure}

The top panel in Fig.~\ref{fig:rhoR} shows the radially-integrated density (areal density) as a function of axial position at the time of peak neutron production, with and without a 50T magnetic field applied. The average areal density is lower in the magnetized case, as the significant magnetic pressure decreases the compression. It can be seen that the size of the axial areal density perturbation is decreased with the magnetic field applied, as the magnetic tension stabilizes the perturbation \cite{chandrasekhar1962}. 

In the azimuthal direction, however, the perturbation size is enhanced by magnetization (see bottom panel in Fig.~\ref{fig:rhoR}). In this orientation the magnetic tension does nothing to suppress the perturbation growth. Instead, suppression of electron thermal conduction decreases the stability, as thermal ablative stabilization is reduced; there is less ablation of cold plasma into the hot-spot to stabilize the growth \cite{walsh2019}. This process is still occurring in the axial direction, but the magnetic tension stabilization dominates.

The effect of an applied magnetic field on the perturbation growth is a balance between the magnetic tension stabilizing axial perturbations and the decreased thermal ablative stabilization in all directions \cite{walsh2019}. The balance between the two processes is highly dependent on magnetic field strength and the perturbation wavelength. Shorter wavelengths and higher field strengths are thought to be dominated by magnetic tension, while intermediate wavelengths and smaller magnetic fields are mostly affected by suppression of thermal conduction \cite{walsh2019}. If an artificial perturbation is applied to these cylindrical implosions (for example through contouring of the shell) then these predictions could be tested experimentally, which would be of great benefit to the magnetized ICF community. 

\begin{table*}[t]
  \centering
  \begin{tabular}{llll}
   \textbf{Description} & \textbf{Symbol} & \textbf{Definition} & \textbf{$\Gamma = 1$ Approximation} \\
    Frozen-in-flow factor & $\Gamma$ &  $\frac{|\underline{B}|}{ B_0} \frac{\rho_0}{\rho}$ & 1 \\
    Field Compression Factor & $\Lambda$ & $\frac{|\underline{B}|}{ B_0}$ & $\frac{\rho}{\rho_0}$\\
    Electron Hall Parameter&$\omega_e \tau_e$ & $\propto \frac{| \underline{B}| T_{e}^{3/2}}{n_e}$ &  $\propto \frac{B_0}{\rho_0} T_{e}^{3/2}$\\ 
    Ratio of Thermal to Magnetic Pressure&$\beta$&$\frac{2(P_e + P_i)\mu_0}{|\underline{B}|^2} $ &$\frac{2 (1+Z)\rho_0^2 T_e \mu_0}{m_i B_0^2 \rho}$  \\
    %Resistive Diffusion Timescale&$t_{diff}$&$\frac{R^2 }{\eta}$ & \\
    Ohmic Dissipation Factor&$\Pi$&$\frac{t_{\mathrm{stag}}}{t_{\mathrm{diff}}} \frac{1}{\beta}$ & $\frac{B_0}{\rho_0 ^3} \frac{t_{\mathrm{stag}} \rho}{\tau_{ei}T_e} \alpha_{\perp}^c \frac{m_i^2 m_e Z}{2(Z+1) e}$\\
    Maximum Unstable Magnetic Tension Mode&$ L_{hs}/\lambda_B$& $\frac{\mu_0 (\rho_h - \rho_l) g L_{hs}}{|B^2|}$ &$\frac{L_{hs} \rho_0^2 \mu_0 (\rho_{\mathrm{shell}} - \rho) V_{\mathrm{imp}}}{\rho^2 B_0^2 t_{\mathrm{stag}}}$\\
    Twisting Factor&$t_{\mathrm{stag}}/t_{\mathrm{twist}}$& $\frac{2 |\underline{j}| t_{\mathrm{stag}}}{L_{hs} e n_e}$ &$\frac{2}{L_{hs}} \frac{B_0}{ R_0}  \frac{\rho^{1/2}}{\rho_0^{3/2}} \frac{m_i}{\mu_0 Ze}$\\
    Ratio of Thermal to Current-Driven Transport&$\Xi$&$\frac{\gamma_{\perp}^c \tau_{ei} T_e e n_e \mu_0}{|\underline{B}|m_e}$&$\frac{\gamma_{\perp}^c \tau_{ei} T_e e  \mu_0 \rho_0 Z}{m_i m_e B_0}$

  \end{tabular}
  \caption{Non-dimensional numbers determining the relative importance of different physical processes in magnetized cylindrical implosions. These factors can be used to optimize an experiment to explore specific phenomena (see Sec.~\ref{sec:design}) or can be inferred by using the frozen-in-flow approximation along with spectroscopy (see Sec.~\ref{sec:diagnosis}) }
  \label{tab:1}
\end{table*}

The significance of the magnetic tension stabilizing perturbations can be quantified by using linear magnetized Rayleigh-Taylor theory \cite{chandrasekhar1962}. The minimum unstable length-scale due to the magnetic field alone (i.e. neglecting thermal conduction) can be written as:
\begin{equation}
    \lambda_B = \frac{|B^2|}{\mu_0 (\rho_h - \rho_l) g}, \label{eq:lambda_B}
\end{equation}
where $g$ is the acceleration of an unstable interface between a heavy and light fluids of densities $\rho_h$ and $\rho_l$ respectively. For a stagnating hot-spot the light fluid is the hot-spot density $\rho$, while the heavy fluid is the peak density $\rho_{\mathrm{shell}}$. Again, by assuming frozen-in-flow, the minimum unstable wavelength due to magnetic tension can be written in terms of initial experiment conditions and stagnation quantities:
\begin{equation}
    \lambda_B = \frac{\rho^2 B_0^2 t_{\mathrm{stag}}}{\rho_0^2 \mu_0 (\rho_{\mathrm{shell}} - \rho) V_{\mathrm{imp}}},\quad \text{if } \Gamma = 1. \label{eq:lambda_B_rho0}
\end{equation}

All the key parameters for quantifying the relevance of different magnetized processes are summarized in Table~\ref{tab:1}; namely, \emph{(a)} $\Gamma$ determines the effectiveness of magnetic flux compressing with the plasma, \emph{(b)} $\Lambda$ is a measure of how compressed the magnetic field becomes, \emph{(c)} the magnetization of the electron orbits is captured by $\omega_e \tau_e$, \emph{(d)} the importance of magnetic pressure to the plasma dynamics is characterized by $\beta$, \emph{(e)} $\Pi$ determines the significance of Ohmic dissipation, \emph{(f)} $L_{hs}/\lambda_B$ is the axial mode number above which magnetic tension stabilizes the hot-spot to linear Rayleigh-Taylor growth, \emph{(g)} the twisting factor $t_{\mathrm{stag}}/t_{\mathrm{twist}}$ determines whether the Hall term is expected to cause significant twists in the magnetic field profile, and finally, \emph{(h)} $\Xi$ then determines if thermal or current-driven transport is more important.

\section{Magnetization Regimes in Cylindrical Implosions\label{sec:design}}

In this section, modifications to the nominal experiment are explored through the non-dimensional factors in Table \ref{tab:1}. It is shown that the metrics can be used to design implosions for different regimes where certain processes dominate magnetic flux compression, thermal transport and bulk hydrodynamics.

Throughout this section, metrics are taken as their definition and do not utilize the frozen-in-flow ($\Gamma = 1$) approximation. The 2D simulations are reduced to 0D metrics by taking the burn-averaged values.

%% FIG. 5 - FROZEN IN FLOW FACTOR
\begin{figure}
	\centering
	\includegraphics[scale=0.73]{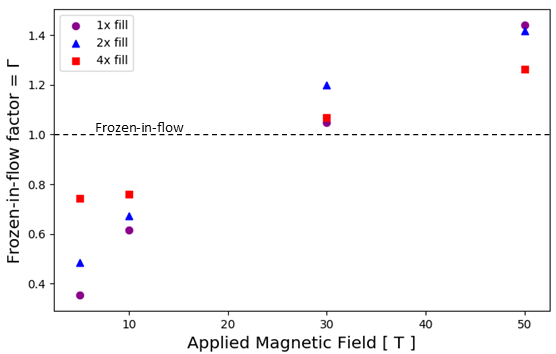}\caption{The frozen-in-flow factor  $\Gamma = |\underline{B}| \rho_0 / B_0 \rho$ for 2D simulations over a range of applied magnetic field strengths and initial gas fill pressures. $\Gamma = 1$ means that the magnetic flux has perfectly compressed with the plasma density. Low applied field strengths result in extra Nernst de-magnetization of the hotspot. $\Gamma>1$ can be obtained due to 2D motion reducing the plasma compression while the magnetic flux continues to compress. \label{fig:FCF}}
\end{figure}

First, the topic of magnetic flux compression is explored. Fig.~\ref{fig:FCF} plots $\Gamma$ for a series of 2D simulations with various fill densities and applied field strengths. Note that varying the fuel density is used to modify the stagnation conditions, with all densities quoted as a multiple of the nominal deuterium fill density of 1.81 kg/m$^3$. While higher densities may not be practical from a target fabrication standpoint, lowering the fill density increases the implosion convergence ratio to an extent that is difficult to resolve with current simulations. Increased convergence may also be undesirable for the onset of hydrodynamic instabilities. Higher fill densities could be achieved by using a foam fuel instead. However, the current study keeps the fill composition as $D_2$ in order to differentiate the impact of changing ionization state from changing fill density. The calculation of $\Gamma$ uses the burn-averaged density and magnetic field strength.

As shown in Fig.~\ref{fig:FCF}, low applied field strengths result in the flux compression being far below the frozen-in-flow approximation. This is primarily due to the Nernst term, which transports magnetic field in the same direction as the electron heat-flow. Like the heat-flow, the Nernst term is suppressed at higher magnetizations. The lower fill density cases have larger contributions from Nernst, which is due to a combination of the hot-spot being hotter and having a smaller radius, both amplifying the temperature gradients. 

For higher applied fields, the Nernst term is almost completely suppressed. At 50~T, $\Gamma > 1$ for all fill densities simulated, meaning that the magnetic field has compressed more than the plasma itself. At first this seems counter-intuitive, as Nernst moves magnetic field out of the hot-spot and resistive diffusion acts to lower the field strength. However, here it is actually found that resistive diffusion can increase the field strength in the hot-spot and that axial motion allows for additional magnetic flux compression, as explained hereafter. It is important to remember that the compressed field and density values used to calculate $\Gamma$ plotted in Fig.~\ref{fig:FCF} are burn-averaged quantities, which means they sample the hot core at the center more than the hot-spot edge. However, the hot-spot is lower density than the core periphery, meaning that the compressed field in the former is lower than in the surrounding plasma. Therefore, resistive diffusion acts to move magnetic flux from the compressed low temperature fuel to the low density hot-spot, increasing the plasma magnetization in the process.

The axial plasma motion also enhances the magnetic flux compression. As this is a cylindrical implosion, there are significant end losses out of the hot-spot \cite{hansen2020}. These unconfined hot-spots can be compressed to a smaller radius. While the axial motion removes mass from the hot-spot, it does not remove magnetic flux, which is predominantly axial itself. Therefore, more axial motion of plasma allows higher compression of the magnetic field. In theory, the greatest magnetic flux compression can be reached by driving extreme axial flows. If the aim of the experiment is to reach large electron magnetizations, however, large axial flows result in lower hot-spot temperatures and low burn durations. The low fill 50T simulation has the largest $\Gamma$ of greater than 1.4, as the axial flows driven are larger than the higher fill densities. 

Note that $\Gamma < 1$ is always expected in experiments where the fuel is laser preheated, such as MagLIF \cite{gomez2020}. The fuel becomes hot even before the magnetic flux is compressed, resulting in low electron magnetizations and large temperature gradients, allowing Nernst demagnetization to dominate.

%% FIG. 6 - BURN AVERAGED B/B_0
\begin{figure}
	\centering
	\includegraphics[scale=0.7]{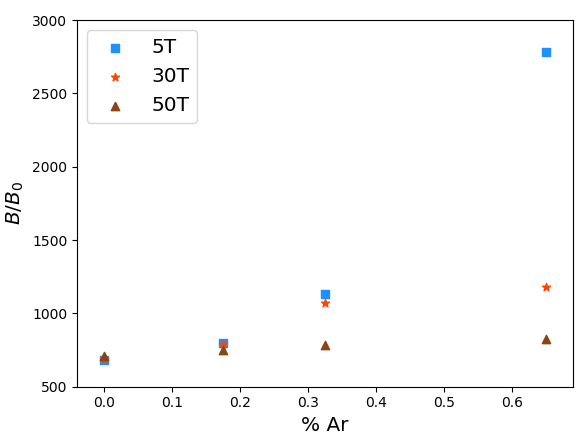}\caption{Burn-averaged field compression factor $\Lambda=|\underline{B}|/B_0$ for 2D simulations over a range of argon dopant concentrations and with applied magnetic fields of 5~T, 30~T and 50~T. Lower applied field strengths are easier to compress, as electron heat-flow magnetization results in a high temperature, low density hot-spot. In addition, large magnetic pressures result in lower compression. Introducing a dopant to the fuel increases magnetic field compression. For a large initial field strength, the magnetic pressure sets the magnetic field compression (rather than the plasma thermal pressure), reducing the compression factor dependence on dopant concentration. \label{fig:B_B0} }
\end{figure}

Mid- or high-Z dopant mixed into the capsule fuel helps to increase magnetic field compression. In particular, Fig.~\ref{fig:B_B0} plots the compression factor $\Lambda$ as a function of argon concentration --the percentage quoted is per deuterium atom--  for 2D simulations with 5~T, 30~T and 50~T applied fields. While applying a 50~T field invariably results in higher compressed field strengths than a 30~T field, the compression factor is always lower. This is because the higher applied field strengths result in lower hot-spot compression. The reasons for this are twofold. Firstly, suppression of thermal conduction reduces the hot-spot compressibility, increasing the temperature at the cost of the density (giving the same stagnated pressure) \cite{betti2001}. Secondly, the final total pressure is the same; higher magnetic pressure results in a lower thermal pressure \cite{hansen2020}. 

\begin{figure}
\centering
		\begin{subfigure}[b]{0.4\textwidth}
			\centering
			\includegraphics[scale=0.73]{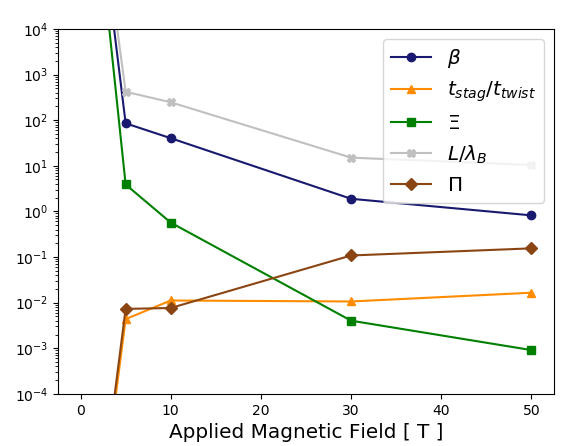}
			\caption{}
			\label{fig:Press_balance}
		\end{subfigure}
		\begin{subfigure}[b]{0.4\textwidth}
			\centering
			\includegraphics[scale=0.73]{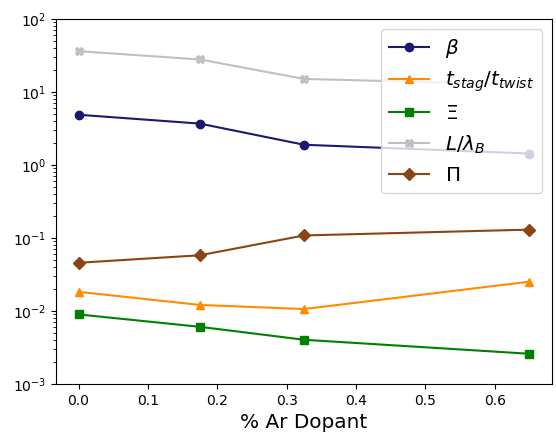}
			\caption{}
			\label{fig:Press_balance_Ar}
		\end{subfigure}
		
		\caption{Magnetization metrics from Table \ref{tab:1} for: (a) varying the applied field strength while keeping the argon concentration at 0.325\% by atom; (b) varying the argon fuel dopant concentration while keeping the applied field strength at 30~T. }
		\label{fig:Press_balance_both}
	\end{figure}

For the 5~T case in Fig.~\ref{fig:B_B0}, increasing the dopant concentration greatly increases the magnetic flux compression. This is because the hot-spot radiatively cools, reducing the thermal pressure and allowing for more compression. The same is true in principle for the 50~T case. However, the hot-spot deceleration is not just dominated by the thermal pressure when 50T is applied, since the magnetic pressure is significant --as will be seen later in Fig.~\ref{fig:Press_balance}, the ratio of thermal to magnetic pressure is $\sim1$. Therefore, the dopant has less of an impact on the magnetic field compression and leads to the magnetic field compression factor being relatively independent of the dopant concentration, as seen in Fig. \ref{fig:B_B0}. 

\begin{figure}
	\centering
	\includegraphics[scale=0.65]{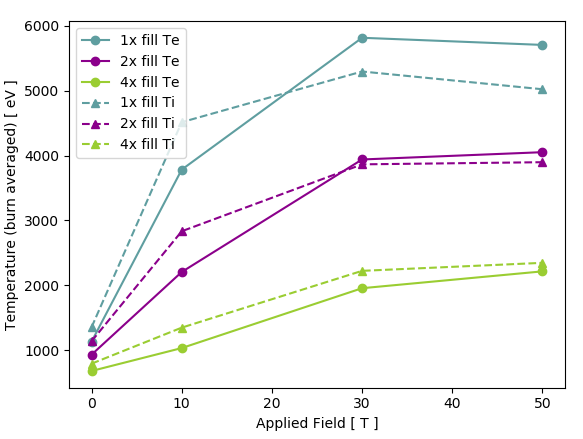}\caption{Burn averaged electron and ion temperatures for 2D simulations for a range of gas fill pressures and applied field strengths. Larger magnetic fields reduce the electron thermal conduction losses from the hot-spot, raising the temperature. Higher fill pressures result in lower plasma magnetizations, reducing the temperature enhancement. In the regime of large magnetic pressure, the electron temperature can exceed the ion temperature due to Ohmic dissipation. \label{fig:Ti_Te}}
\end{figure}

Figure \ref{fig:Ti_Te} shows the burn-averaged electron and ion temperatures of 2D simulations scanning both applied field strength and fill density. For these cases, a dopant concentration of 0.325\% is used. As the magnetic field strength is increased, the thermal conduction from the hot-spot into the cold fuel is suppressed, raising the core temperature. For the 50~T low fill density case the electron Hall parameter reaches a burn-averaged value of over 300, reducing the thermal conductivity perpendicular to the magnetic field to $\kappa_{\bot}/\kappa_{\parallel} = 5\times 10^{-5}$.

The higher fill densities reach lower temperatures, reducing the electron magnetization. Therefore, the enhancement in temperature is lower for a higher initial fill density. 

An interesting feature in Fig.~\ref{fig:Ti_Te} is the electron temperature being larger than the ion temperature in the highly magnetized low fill density cases. This is due to the Ohmic dissipation converting magnetic energy into electron thermal energy. In unmagnetized implosions the electron population loses energy faster than the ions due to their additional mobility. Typically it is through electron-ion collisions that the ion population loses its energy, but in the magnetized case a significant proportion of the implosion kinetic energy is used to compress the magnetic field ($\beta \approx 1$). Once the field starts to diffuse and relax, this potential energy is transferred to the electrons, which are resistively heated. For $B_0 = \unit[50]{T}$, the Ohmic heating factor is $\Pi = 0.13$ at the hot-spot edge, indicating that a significant proportion of the magnetic energy is being transferred into electron thermal energy. 

Measuring a higher electron temperature than ion temperature would be a definitive observation of this process taking place. However, there may be complications due to the ion temperature typically being inferred by the spectra of fusion-produced neutrons \cite{appelbe2014}. Doppler shifting of the spectra by fluid motion can increase the apparent ion temperature, which may be comparable in size to the $\approx \,500~\unit{eV}$ difference in electron and ion temperatures reported here. 

Figure \ref{fig:Press_balance} displays the dependence of a range of metrics from Table \ref{tab:1} on the applied field strength. Universally, these all show the importance of the induced current increasing with applied field strength. For an applied field of 50~T the hot-spot magnetic pressure is approximately the same size as the thermal pressure, giving a $\beta \approx 1$. Along with this, the axial mode numbers that are suppressed by magnetic tension ($L_{hs}/\lambda_B$) decreases to 12, meaning that all high mode features are expected to be suppressed in the axial direction. The importance of thermal transport of magnetic field and electron energy is almost completely suppressed by electron magnetization, which can be seen from $\Xi < 0.001$. While the magnetic field twisting factor is still small for  a seed B-field of 50~T, this could be enhanced by reducing the driving length of the target. 

Figure \ref{fig:Press_balance_Ar} then shows how the induced-current metrics vary with the argon dopant percentage for an applied magnetic field of 30~T. Again, the added dopant results in additional radiative losses, allowing the hot-spot to become more compressed. Similarly, the magnetic field compresses further and gives larger induced currents. As a result, the magnetic pressure is greater, the Ohmic heating becomes more important, the magnetic tension stabilizes more wavelengths, the importance of thermal transport is reduced and the Hall term twists the magnetic field more effectively.

% SECTION VI
\section{Argon K-shell spectroscopy \label{sec:diagnosis}}
\begin{figure}
\centering
\begin{subfigure}[b]{0.45\textwidth}
\centering
\includegraphics[width=\textwidth]{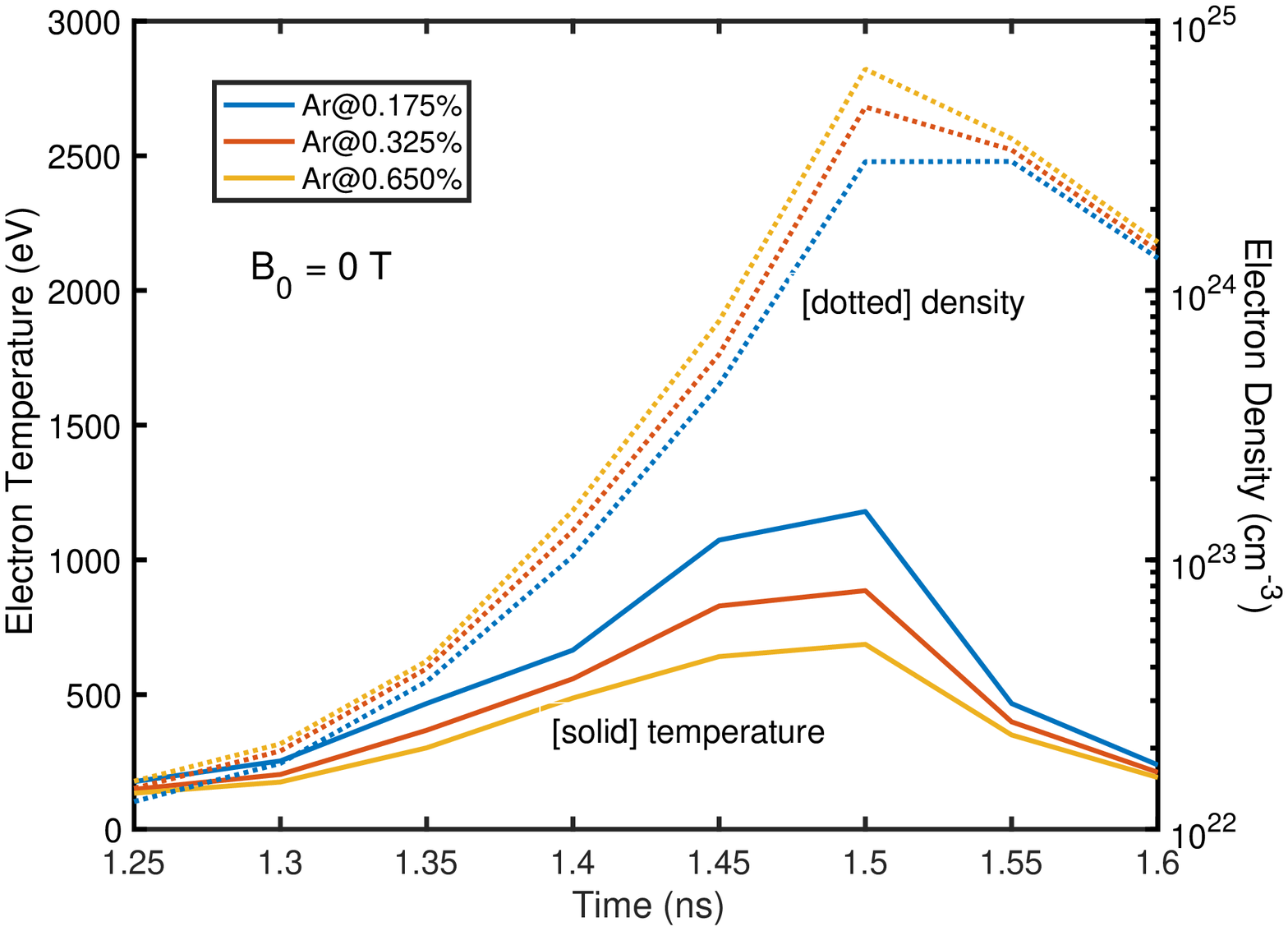}
%\caption{}
%\label{fig:Press_balance}
\end{subfigure}
\begin{subfigure}[b]{0.45\textwidth}
%\centering
\includegraphics[width=\textwidth]{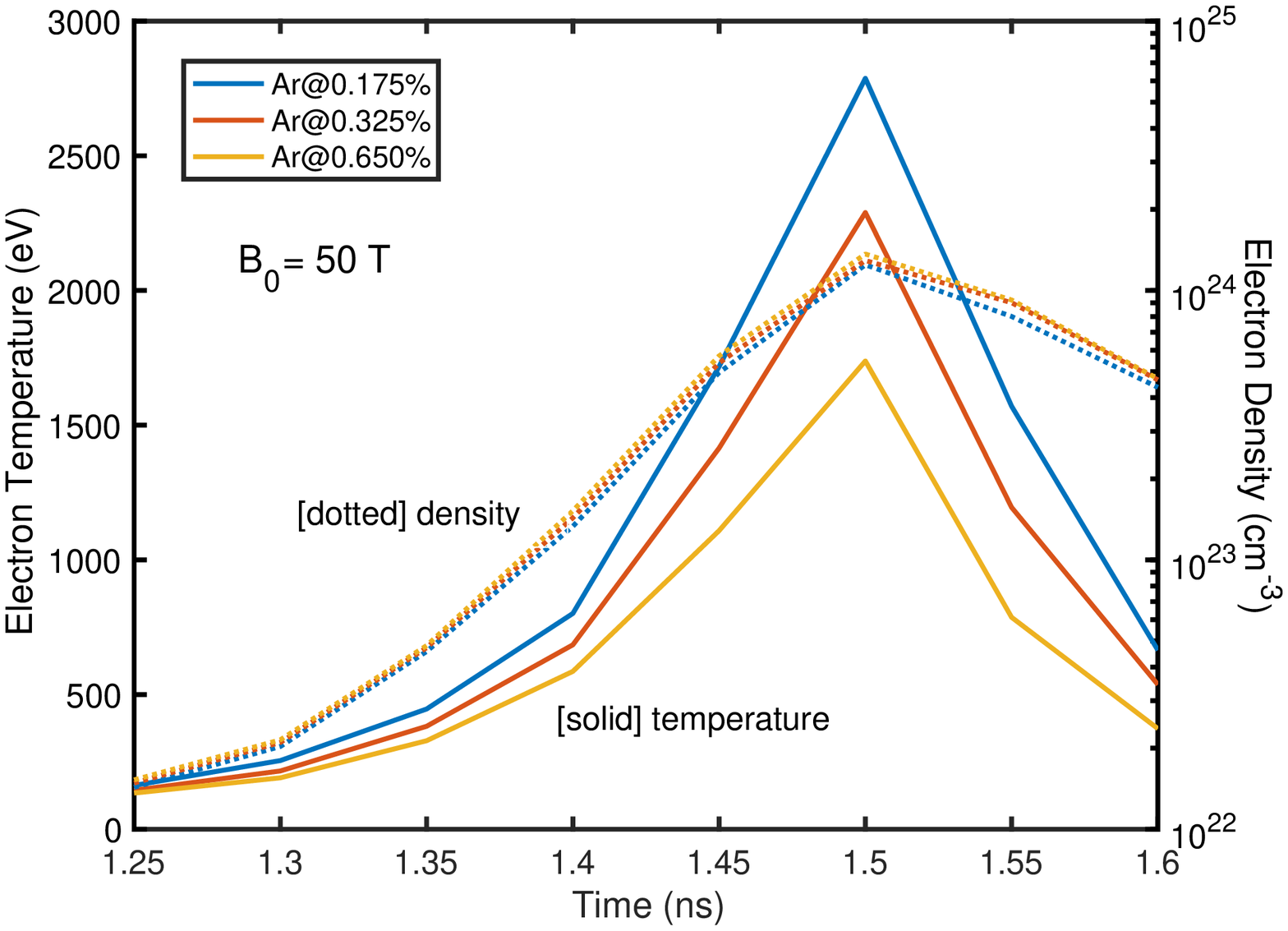}
%\caption{}
%\label{fig:Press_balance}
\end{subfigure}

%\begin{subfigure}[b]{0.45\textwidth}
%%\centering
%\includegraphics[width=\textwidth]{./Figures/B_hist_50T.eps}
%%\caption{}
%%\label{fig:Press_balance}
%\end{subfigure}

\caption{Time history of core conditions from 2D Gorgon simulations of cylindrical implosions as a function of different argon dopant concentrations. Shown values are obtained as mass-weighted averages within the radial core boundary and over the central \unit[400]{\micro m}-long portion of the imploded cylinder. Top and bottom panels show electron temperature and electron density from non-magnetized and magnetized cases, respectively.}
%The bottom panel shows the time history of the magnetic field for an initial seed B-field of 50 T.}
\label{fig:time_histories_mw}
\end{figure}

Here we address the application of X-ray spectroscopy for the characterization of core conditions. The Gorgon MHD simulations show that the compression of the B-field changes the hydrodynamic behavior, thus producing a variety of magnetization phenomena whose underlying physics demand a deeper understanding as discussed in Sec.~\ref{sec:New_Physics} and \ref{sec:design}. Since changes in the hydrodynamics translate into significant variations in the radial profiles of core temperature and density throughout the implosion collapse compared to the unmagnetized case, a way to bring information about the associated physics is by diagnosing the conditions of the imploding core. In this regard, given the success in past experimental campaigns of spherical implosions at OMEGA~\cite{regan2002,welser2007,florido2011,florido2014,nagayama2014} and also in laser-heated cylindrical plasma experiments performed at Z related to the MagLIF preheat stage~\cite{carpenter2020a,carpenter2020b}, here we propose to rely on time-resolved Ar K-shell spectroscopy. This diagnostic exploits two basic properties of the Ar K-shell spectrum emitted from hot, dense plasmas: (1) the Stark-broadened line shapes strongly depend on density and are relatively insensitive to variations in electron temperature, (2) the relative intensity distribution of K-shell lines and their associated satellites are sensitive to variations in electron temperature and density through the dependence to these parameters on the atomic level population kinetics.  The Ar K-shell spectroscopic technique works properly for plasma electron densities and temperatures typically in the ranges of $n_e\sim 10^{23}-10^{24}$~cm$^{-3}$ and $T_e\sim 600-2000$~eV, since for these conditions the Stark effect is the dominant line broadening mechanism, and He-like and H-like Ar are significantly populated so that their corresponding line emission becomes bright and useful for analysis. In this regard, Fig.~\ref{fig:time_histories_mw} shows the time histories of mass-weighted average electron temperature and electron density of the imploded core for different argon concentrations obtained from 2D Gorgon simulations. Overall, average core conditions for the non-magnetized implosions fit well within the ranges above, while for the magnetized case the temperature can go above 2~keV around stagnation. Still, since the new physics and main goal here focuses on exploring and providing evidence on extreme magnetization phenomena, we consider appropriate to use the well-established Ar K-shell spectroscopic technique and perform an assessment of its potential application, in order to find out how far the technique can be pushed for the characterization of core conditions in the magnetized scenario. It is also worth noting that despite the high B-field values achieved in the compressed core ($> 10$~kT around stagnation), Stark-Zeeman spectroscopy cannot be used to obtain an estimation of compressed B-field in an experimental scenario. Given the range of electron densities achieved, we performed systematic calculations of Stark-Zeeman line profiles based on molecular dynamics simulations~\cite{gigosos2018} and checked that the broadening of Ar K-shell lines produced by the Stark effect blurs the characteristic Zeeman pattern even in the case of the expected achievable B-field. Zeeman features will be further washed out by radiation transport effects through the imploded core and instrumental broadening. Details of these calculations are beyond of the scope of this paper and will be addressed in a forthcoming publication.

%%%%%%%%% FIGURE 9 %%%%%%%%%%%%%
\begin{figure*}
\centering
\includegraphics[width=\textwidth]{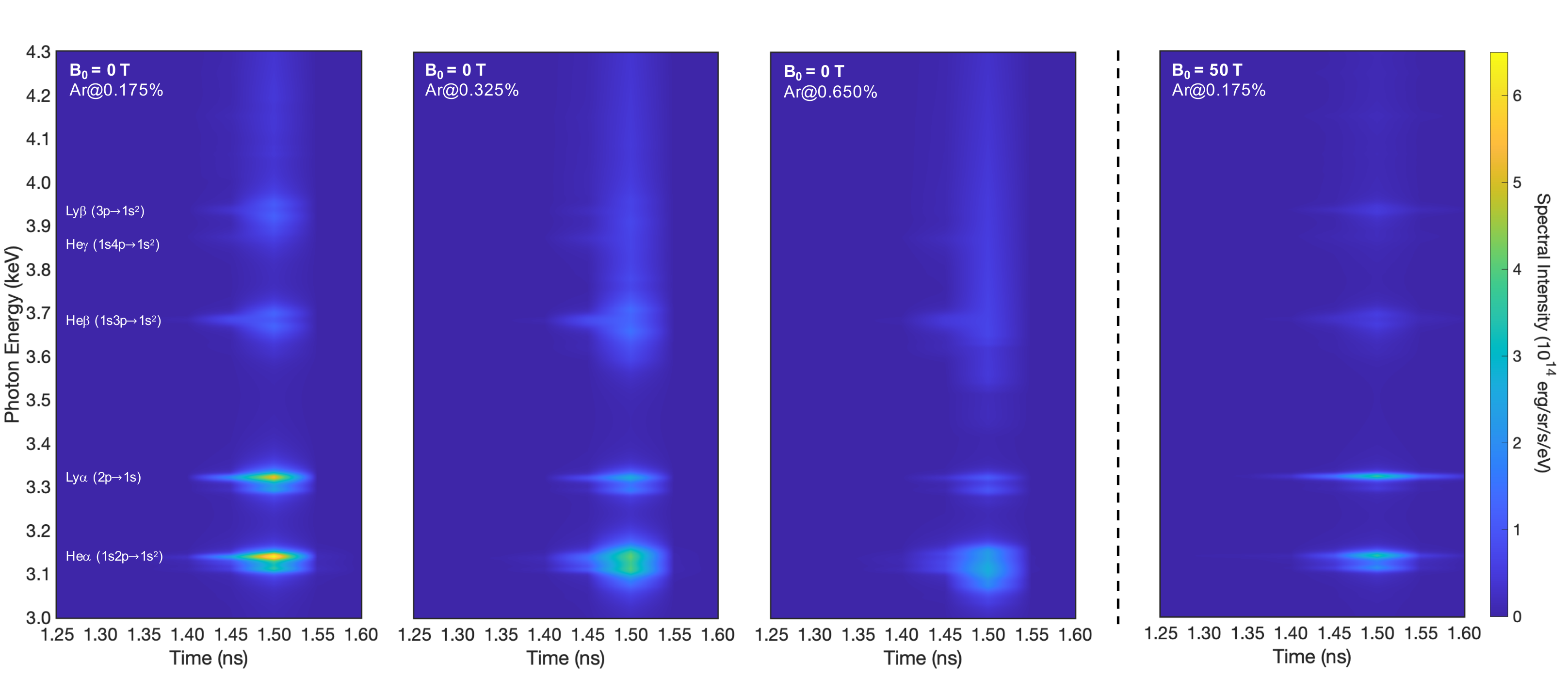}
\caption{Comparison of synthetic images of Ar K-shell time-resolved spectra for unmagnetized and magnetized ($B_0=50$~T) implosions. For the non-magnetized case, changes in the spectra due to different Ar concentrations are shown. For the magnetized case, the sensitivity with Ar concentration is lost and only the case for Ar at 0.175\% is displayed --see the text for details--.}
\label{fig:streaked_images}
\end{figure*}

In an effort to have a realistic representation of the time-resolved Ar K-shell spectra to be measured in a real experiment, we have post-processed the 2D MHD Gorgon simulations and performed detailed atomic-kinetics and radiation transport calculations throughout the evolution of the imploding cylinder. Line of sight is assumed to be perpendicular to the cylinder axis, so that to compute the emergent intensity the radiation transport equation must be solved along chords inside the plasma source parallel to the line of sight. The solution to this problem can be well approximated by a system of effective slabs~\cite{joshiphd} resulting from the discretization of the cylindrical plasma in a finite number of cylindrical shells, which here was chosen to exactly match the radial grid used in the Gorgon MHD simulations. Temperature and density conditions of each cylindrical shell are taken as the spatial average for a given time over the central \unit[400]{\micro m}-long portion of the imploded core along the axial direction. This is a reasonable simplification, since according to MHD simulations --see Fig.~\ref{fig:rho_Te_2D}--, conditions of the imploded cylinder show little dependence on the axial direction in the central region. The required emissivities, opacities and atomic level population distributions were calculated with the collisional-radiative model ABAKO~\cite{florido2009} for each pair of core electron temperature and density over the corresponding radial grid as obtained from the Gorgon MHD output, i.e. no interpolation over pre-calculated emissivity and opacity databases was applied. Bound-bound, bound-free and free-free contributions from the Ar-doped deuterium plasma within the spectral range of interest were included. In particular, relevant bound-free cross sections~\cite{florido2010} were computed with the LANL suite of codes~\cite{fontes2015}. Energy levels and line transitions rates for the required Ar ion stages were computed with the atomic structure code FAC~\cite{gu2008}. For given plasma conditions, ABAKO explicitly includes all non-autoionizing and autoionizing states consistent with the corresponding ionization potential depression, which is estimated according to the Stewart-Pyatt model~\cite{stewart1966}, modified to account for the argon-deuterium mixture. An escape factor model was used to account for radiation trapping effects on the atomic-level population kinetics~\cite{mancini1987}. Importantly for this application, detailed Stark-broadened line shapes for Ar He$\alpha$, He$\beta$, He$\gamma$, Ly$\alpha$, Ly$\beta$, and Ly$\gamma$, and associated satellites were calculated with the MERL line-shape code~\cite{mancini1991}, including ion-dynamics effects~\cite{boercker1987}. Recently, MERL Stark-broadened line shapes of Ar K-shell have been further validated for spectroscopic diagnosis applications against full molecular dynamics simulations~\cite{gigosos2021}. 

Figure~\ref{fig:streaked_images} shows the time evolution of the computed Ar K-shell emission, i.e.~synthetic \textit{streaked} images, for unmagnetized and magnetized cases assuming a seed B-field of $50$~T. Besides Doppler and Stark broadening mechanisms, synthetic spectra also accounts for instrumental broadening. Consistently with available streaked spectrometers on OMEGA for the photon energy range of interest, a spectral resolution of $E /\Delta E\approx 500$ was assumed. Main line transitions in He-like and H-like ions are labeled, i.e. He$\alpha$~(1s2p$\rightarrow$1s$^2$), He$\beta$~(1s3p$\rightarrow$1s$^2$) and He$\gamma$~(1s4p$\rightarrow$1s$^2$) in He-like Ar, and Ly$\alpha$~(2p$\rightarrow$1s), Ly$\beta$~(3p$\rightarrow$1s) and Ly$\gamma$~(4p$\rightarrow$1s) in H-like Ar. There are significant differences between the expected time-resolved spectra of a non-magnetized implosion and a magnetized one. Moreover, the metrics analysis of Sec.~\ref{sec:design} demonstrated that a extreme magnetization regime is already attainable for a seed B-field of 30~T, thus also expecting noticeable differences --similar to those shown in Fig.~\ref{fig:streaked_images}-- in the spectral output when compared to the unmagnetized case. Therefore, based on our simulations, in a potential experimental realization the observation of the time-resolved Ar K-shell emission should be able to distinguish between the unmagnetized and magnetized scenarios (at least for $B_0>30$~T). A detailed analysis of spectroscopic data would bring further information about the magnetization state, as pointed out below.   

For interpretation, it is useful to correlate the synthetic streaked images in Fig.~\ref{fig:streaked_images} with the time histories of electron temperature and density shown in Fig.~\ref{fig:time_histories_mw}. In the unmagnetized case, line emission turns on at $t\gtrsim1.35$~ns and lasts for $\lesssim200$~ps. Figure~\ref{fig:streaked_images} shows that, overall, as expected, signal intensity increases with time until stagnation ($t=1.5$~ns), when the average temperature and density reach their maximum values, and decreases after. Lines are narrower early in time, then gradually broaden until stagnation time and narrowing down afterwards, mainly due to Stark-broadening which increases with increasing electron density. When comparing the results for different Ar concentrations, the ratio of He-like to H-like for the same type of transition increases with the Ar amount. This is a result of radiative cooling, i.e. core temperature decreases and He-like ion population is favored relative to that of H-like ion. Also, core compressibility is higher for the case of Ar at 0.65\%, so we observe broader lines when compared to the results for Ar at 0.175\%. For Ar at 0.65\% the overall spectral intensity drops compared to Ar at 0.175\%, which, despite the density increasing, can be explained because the plasma cooling remains as the dominant effect; i.e. for Ar at 0.65\%, the core average temperature stays below $700$~eV throughout the implosion and, relative to the case of Ar at 0.175\%, the ion population distribution shifts to lower ionization stages, so that the number of K-shell ion emitters decreases accordingly.  

\begin{figure}
\centering
\includegraphics[width=0.45\textwidth]{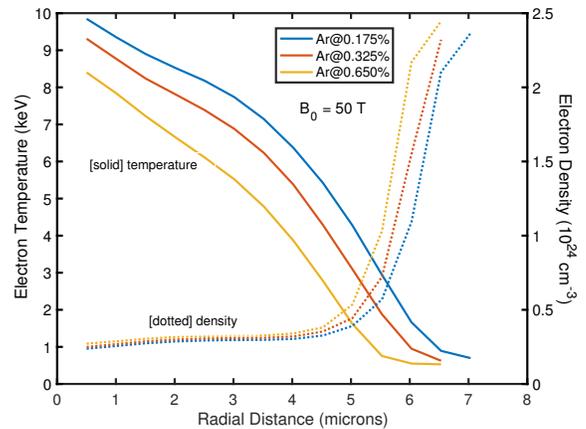}
\caption{Radial profiles of core conditions at stagnation time ($t=1.5$~ns) for a magnetized cylindrical implosion (with $B_0=50$~T). Results for different Ar concentrations are shown.}
\label{fig:stag_radial_profiles}
\end{figure}

In the magnetized case, line emission lasts longer ($\sim250$~ps) and the overall signal intensity is lower than compared to a non-magnetized implosion. Moreover, since the effect of magnetic pressure reduces the core compressibility, electron density and therefore the impact of Stark-broadening decrease, which ultimately leads to narrower lines compared to the unmagnetized scenario. Interestingly, contrary to what was found for unmagnetized implosions, streaked images of K-shell emission show little sensitivity to the Ar concentration --only the case for Ar at 0.175\% is displayed in Fig.~\ref{fig:streaked_images} due to this reason--. This is due to the large gradients in core conditions along the radial direction throughout the collapse of the implosion for the magnetized case. For illustration, Fig.~\ref{fig:stag_radial_profiles} shows the radial profiles at stagnation time of core electron temperature and density averaged in the axial direction over the central \unit[400]{\micro m}-portion of the imploding cylinder. The temperature radial profiles show that, regardless of the Ar amount, a lower bound of $\sim700$~eV is found at the core periphery. Temperature rises radially inwards to values far above $\sim2$~keV, where Ar ions will be fully ionized. Thus, it is clear that K-shell emission for the magnetized case will be only representative of the core outermost portion. When temperature spatial profiles are correlated with the corresponding  density profiles, one realizes that, no matter the Ar concentration, the radiation transport comes across basically the same ranges and combinations of core conditions relevant for K-shell line emission. Accordingly, since a similar behavior is found over the implosion collapse, the resulting time-resolved spectra show little dependence on the Ar amount. No significant gradients are observed in the radial profiles for the unmagnetized case. In Sec.~\ref{sec:design}, the variation of Ar concentration was suggested as a fine-tuning knob for achieving different magnetization regimes for a given seed B-field value. However, since the sensitivity with Ar concentration is lost, our simulations here show that Ar K-shell spectroscopy would not be useful to distinguish between such regimes. Still, if ideally multiple spectrometers were available in the experimental facility to simultaneously measure different spectral ranges, we could keep using the dopant concentration as a control knob and speculate with doping the targets with small amounts of different dopants, for instance mixing Ar and Kr. Ar K-shell spectroscopy seems to work properly for the characterization of core conditions in the unmagnetized case and, potentially, Kr K-shell and L-shell spectroscopy could provide a better characterization of magnetized implosions, even enabling to retrieve the sensitivity of the emission spectra to the dopant concentration. This subject is under investigation and will be published elsewhere.  

%%% FIGURE 10

\begin{figure}
\centering
\includegraphics[width=0.5\textwidth]{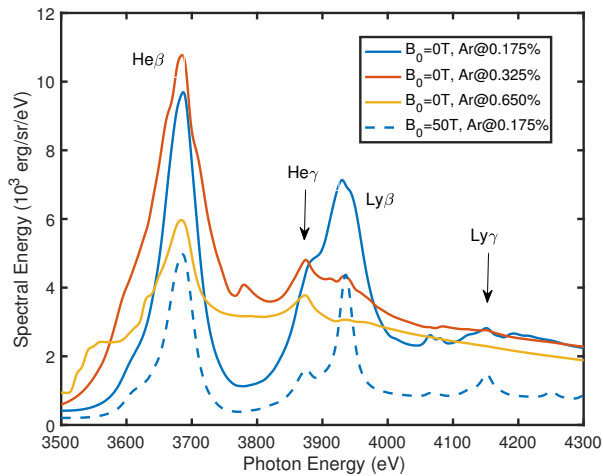}
\caption{Comparison of synthetic Ar K-shell time-integrated spectra for non-magnetized and magnetized ($B_0=50$~T) implosions. For the unmagnetized case, changes in the spectrum due to different Ar concentrations are shown. For the magnetized case, the sensitivity with Ar concentration is lost and only the case for Ar at 0.175\% is displayed --see the text for details.}
\label{fig:time_integrated_spectra}
\end{figure}

Figure~\ref{fig:time_integrated_spectra} shows synthetic time-integrated spectra for the unmagnetized and magnetized cases to evaluate potential data from this diagnostic on OMEGA. Spectra are shown for a photon energy from 3500 to 4300 eV, since temperature and density diagnosis based on K-shell spectroscopy usually relies on the $\beta$ and $\gamma$ lines region of the spectrum --$\alpha$ lines are much more sensitive to spatial gradients and absorption effects than others due to their characteristic large optical depths--. As already discussed, there is little variation with Ar concentration for the magnetized case, so that only the case for Ar at 0.175\% is displayed. The comparison of time-integrated spectra obtained at different Ar concentrations for a non-magnetized implosion allows us to answer the question about the optimum dopant concentration for spectroscopic purposes. For Ar at 0.175\%, strong He$\beta$ and Ly$\beta$ emissions are observed, which is critical for a reliable spectroscopic analysis. For Ar at 0.325\% the Ly$\beta$ emission is almost lost and the situation is even worse for Ar at 0.650\%. The same conclusion is found when doing the comparison for time-resolved spectra. Besides the difference in the absolute signal strength, time-integrated spectra from unmagnetized and magnetized cases mostly differ in the broadening of the lines. The higher compressibility of the unmagnetized implosion ultimately leads to a significant impact of Stark-broadening, which is not so severe in the magnetized case. Regarding the overall signal strength, results for non-magnetized and magnetized implosions differ in a factor of $\sim 2$, and both compare within one order of magnitude with calculations done in the past for the campaign of spherical implosions on OMEGA, where the Ar K-shell emission was successfully recorded. Therefore, for all the reasons above, an argon concentration of 0.175\% seems to be a good choice for application of K-shell spectroscopy for the study of the referred direct-drive cylindrical implosions.        
%%% FIGURE 11

Lastly, it is worth noting that, similarly to previous work, a spectroscopic modelling and analysis of the measured time-resolved Ar K-shell emission would allow the determination of the temperature and density of the imploded core throughout the collapse of the implosion. From these spectroscopic measurements and if the magnetic field is assumed to be frozen, estimations for the magnetization metrics in Table~\ref{tab:1} can be obtained. However, in the magnetized scenario, the spectroscopic diagnosis will be more complicated due to the severe gradients in core conditions. A feasibility assessment of the spectroscopic technique based on synthetic data is on progress and will be published in a forthcoming paper.

\section{\label{sec:Conc} Conclusions}

This paper outlines a cylindrical implosion platform for studying highly magnetized plasmas on the OMEGA laser facility. In the highly magnetized regime the imploding shell decelerates on the magnetic field, converting kinetic energy into magnetic. The highly compressed magnetic field induces large electrical currents in the hot-spot; finite plasma resistivity results in Ohmic dissipation, diffusing the magnetic field and heating the electron population. This is expected to result in a higher temperature in the electron population than in the ions, which may be diagnosable. 

The large induced electrical currents can also cause Hall physics to be important, which has not been reported in laser-driven implosions to date. The calculations here estimate that a typical electron in the hot-spot will undergo a 1/3 rotational drift around the cylindrical axis within the fusion time frame, advecting the magnetic field and electron energy distribution. 

The electrical currents also result in perturbation stabilization in the axial direction, which could be studied by using a shell with applied corrugations. Experiments scanning applied field strength and perturbation wavelength would be of great interest to the magnetized ICF community. 

All of these magnetization processes have been summarized as non-dimensional metrics in Table~\ref{tab:1}.

Care has been taken to optimize future potential experiments to reach peak magnetic field strengths. By using a low gas fill, the magnetic field is more easily compressed. However, as the convergence ratio of the capsule increases, the implosion will be more susceptible to perturbation growth, limiting the minimum fill density. Dopants added to the imploding fuel can also be used to radiatively cool the hot-spot, allowing the magnetic field to be more compressible. More axial motion, too, allows for greater flux compression. 

%Another design direction has been investigated, seeking to reach high electron magnetizations. In this case a low dopant concentration is required, although spectroscopic diagnosis of the plasma requires a minimum signal strength. Lower fill densities result in high temperature and low density hot-spots, where electron magnetizations can be in excess of 400. 

X-ray spectroscopy can be used to diagnose the core electron temperature and density. The core temperature is anticipated to increase through the application of a magnetic field, while the density will decrease. The measurement of lower overall core pressure when a large magnetic field is applied would indicate that the magnetic pressure has become significant. By assuming frozen-in-flow, which has been shown with simulations to be a good approximation in the highly magnetized regime, many of the metrics governing which magnetization phenomena are important can be indirectly inferred.

\section{\label{sec:Appendix} Appendix}

Magnetic flux compression has been measured in previous cylindrical implosions on the OMEGA Laser Facility using proton radiography \cite{gotchev2009,knauer2010}. Here, 2D Gorgon simulations are post-processed to create synthetic proton radiographs to compare with these experiments. Good agreement is found when the Nernst term is included in the simulations. These results build confidence in the simulation results in this publication and also (to a lesser extent) in other non-cylindrical Gorgon simulations where magnetic flux compression is an important process.

There are significant differences between the target dimensions used in the bulk of this paper and the target used for this magnetic flux compression validation. Here the target radius is $\unit[430]{\mu m}$, filled with $3$~atm of $D_2$ gas \cite{knauer2010}. A $5$~T magnetic field was applied, which is low enough to have relatively little contribution from magnetic pressure in the hot-spot. The experiment discussed in this appendix was also carried out before the laser non-uniformity was optimized \cite{hansen2018,hansen2018a}, resulting in a lack of axial hot-spot confinement. While this lowers the electron magnetization, the Gorgon simulations actually predict an enhancement to the magnetic flux compression, because the hot-spot is more compressible.

%%%% FIGURE 12 - FOR APPENDIX
\begin{figure}
		\centering
		\includegraphics[width=\columnwidth]{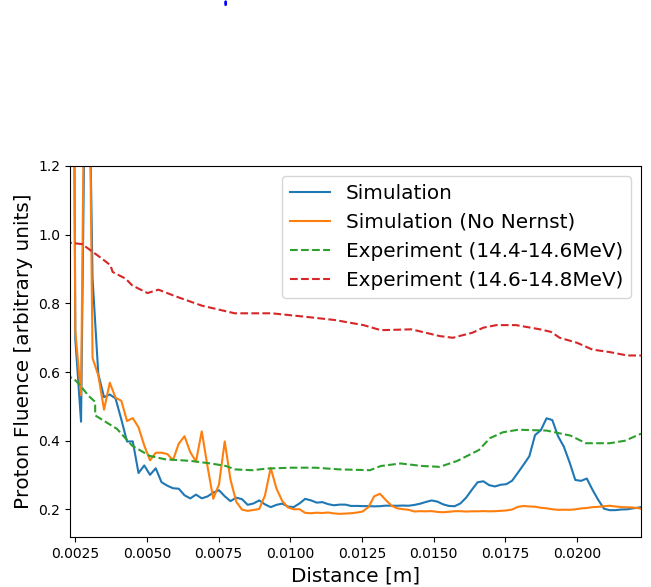}\caption{Proton radiographs of a cylindrical magnetic flux compression experiment. The experimental data is taken from Ref.~\cite{knauer2010}, showing here only the data from the protons that have slowed through the target (energy ranges 14.4-\unit[14.6]{MeV} and 14.6-\unit[14.8]{MeV}). The experimental radiographs show a bump at around \unit[1.8]{cm}, which is consistent with the simulation including Nernst advection. \label{fig:Knauer}}
\end{figure}

The proton source in the experiment was from an exploding pusher, which provides a relatively monoenergetic proton population ($\Delta E/E \sim 0.03$) \cite{knauer2010}. This population is born at \unit[14.7]{MeV}, but is accelerated to \unit[15.2]{MeV} due to charging of the backlighter \cite{knauer2010}. The hot-spot being imaged is small ($\approx \unit[10]{\micro m}$), meaning that very few protons hitting the radiograph are significantly deflected by the compressed magnetic field. On the face of it, this makes imaging the peak field very difficult. However, the protons passing through the compressed plasma are slowed; by looking only at the protons that have had their energies lowered by the compressed plasma, the signal to noise ratio is improved. In the experiment analysis, the proton populations in the energy bands $<\unit[14.4]{MeV}$,$14.4-\unit[14.6]{MeV}$ and $14.6-\unit[14.8]{MeV}$ all show bumps at around $\unit[1.8]{cm}$ deflection \cite{knauer2010}. When the applied field was reversed in direction, the bump moved to $-\unit[1.8]{cm}$, which indicates that the feature is from magnetic field deflections, not from electric field effects. While the signals for each proton range are relatively small compared with the background, the consistency of the feature across several proton energy ranges and across multiple experiments is encouraging.

Synthetic proton radiographs of 2D Gorgon simulations have been plotted in Fig.~\ref{fig:Knauer}. These are produced using a proton tracking code without electric field effects or collisional scattering. A monoenergetic source is assumed. The protons are reported to be emitted over a $\unit[150]{ps}$ burst \cite{knauer2010}, so a Gaussian emission profile with standard deviation of $\unit[75]{ps}$ is used for the synthetic radiographs.

When the Nernst term is included in the simulations a bump is observed in the radiograph at 1.9~cm, similar to that seen experimentally. This bump is not found to be due to the peak field strength reached at the centre of the target, but by the magnetic flux that develops at the hot-spot edge, $\approx \unit[10]{\micro m}$ from the axis. While the field strength at this radius is lower, deflecting the protons less, the total flux of protons into this region is greater, increasing the signal. The magnetic field strength at the hot-spot edge is approximately $4000$~T, which is consistent with the values inferred by the original experiment analysis \cite{gotchev2009}. 

Without the Nernst term included, the synthetic proton radiograph does not compare favourably with the experiment, with a wide spread of proton deflections and no clear bump --see Fig.~\ref{fig:Knauer}. This is the first direct evidence of Nernst in magnetized implosions; Nernst has been indirectly inferred in MagLIF through observation that, above a certain laser preheat energy, the neutron yield plateaus \cite{gomez2020}.  No sensitivity of the radiograph to a Nernst flux limiter in the range 0.04-0.1 was found. While resistive diffusion smooths the radiograph, the cases with and without compare equally favourably to the experiment.

%\note{quote gamma for this Knauer setup}

%\note{Compare figures of Starkzee symmetry at stagnation to Knauer? density and temperature?}

\section{Acknowledgements}
This work performed under the auspices of the U.S. Department of Energy by Lawrence Livermore National Laboratory under Contract DE-AC52-07NA27344. This document was prepared as an account of work sponsored by an agency of the United States government. Neither the United States government nor Lawrence Livermore National Security, LLC, nor any of their employees makes any warranty, expressed or implied, or assumes any legal liability or responsibility for the accuracy, completeness, or usefulness of any information, apparatus, product, or process disclosed, or represents that its use would not infringe privately owned rights. Reference herein to any specific commercial product, process, or service by trade name, trademark, manufacturer, or otherwise does not necessarily constitute or imply its endorsement, recommendation, or favoring by the United States government or Lawrence Livermore National Security, LLC. The views and opinions of authors expressed herein do not necessarily state or reflect those of the United States government or Lawrence Livermore National Security, LLC, and shall not be used for advertising or product endorsement purposes.

This work has also been carried out within the framework of the EUROfusion Consortium and has received funding from the Euratom research and training programme 2014-2018 and 2019-2020 under grant agreement No. 633053. The views and opinions expressed herein do not necessarily reflect those of the European Commission. The involved teams have operated within the framework of the Enabling Research Projects: CfP-AWP17-ENR-IFE-CEA-02 \textit{Towards a universal Stark-Zeeman code for spectroscopic diagnostics and for integration in transport codes} and CfP-FSD-AWP21-ENR-01-CEA-02 \textit{Advancing shock ignition for direct-drive inertial fusion}. 

This material is based upon work supported by the US National Nuclear Security Administration under Award No. DE-NA0003940. The work has also been supported by the Research Grant No. GOB-ESP2019-13 from ULPGC; and by Research Grant No. PID2019-108764RB-I00 from the Spanish Ministry of Science and Innovation.

F.S.-V. acknowledges funding from The Royal Society (UK) through a University Research Fellowship. C.V. acknowledges the support from the LIGHT S\&T Graduate Program (PIA3 Investment for the Future Program, ANR-17-EURE-0027). G. P.-C. acknowledges funding from the French Agence Nationale de la Recherche (ANR-10-IDEX-03-02, ANR-15-CE30-0011). 

\ifdefined\DeclarePrefChars\DeclarePrefChars{'’-}\else\fi

\end{document}